\documentclass[useAMS,usenatbib,usegraphicx]{mn2e}
\title[Gravitational lensing by CDM halos]
      {Gravitational lensing by CDM halos: 
       singular versus nonsingular profiles}
\author[Hugo Martel and Paul R. Shapiro]
       {Hugo Martel\thanks{E-mail: hugo@simplicio.as.utexas.edu} 
        and Paul R. Shapiro\thanks{E-mail: shapiro@astro.as.utexas.edu}\\
Department of Astronomy, University of Texas, Austin, TX 78712}

\begin{document}

\date{Submitted May 5 2003}

\pagerange{\pageref{firstpage}--\pageref{lastpage}} \pubyear{2003}

\maketitle

\label{firstpage}

\begin{abstract}
The gravitational lensing properties of cosmological halos depend
upon the mass distribution within each halo. The description
of halos as nonsingular, truncated isothermal spheres, a 
particular solution of
the isothermal Lane-Emden equation (suitably modified for $\Lambda\neq0$),
has proved to be a useful approximation for the halos which form from
realistic initial conditions in a CDM universe. 
The nonsingular TIS model reproduces many of the quantitative features of
the N-body results
for CDM halos, except in the very center, where CDM N-body
halos show density profiles which vary as
$\rho\sim r^{-\alpha}$, $\alpha\ga1$, instead of a small flat core.
Possible discrepancies between these cuspy halo predictions of the CDM
N-body simulations and observations of the inner mass profiles of dwarf and
LSB disk galaxies based upon their rotation curves and of clusters
based upon strong lensing measurements have led to a search for other
diagnostics. A description of the lensing by TIS halos would be
useful in this regard, as a self-consistent model for CDM halos in a proper
cosmological context, nonsingular but otherwise consistent with the
CDM N-body results.

We derive here the basic lensing properties of individual TIS halos.
For comparison, we also consider three
singular profiles: the Navarro-Frenk-White density profile, the singular
isothermal sphere, and the Schwarzschild lens. For all profiles, we compute
the critical curves and caustics, the image separation, the magnification
and brightness ratio, the shear, the time delay, and the average shear
inside the tidal radius. This provides tools for studying the statistical
properties of lensing by TIS and other lenses
in the context of a theory of cosmological structure formation.
\end{abstract}


\section{INTRODUCTION}

The gravitational lensing of distant sources has in recent years become
one of the most powerful tools in observational cosmology (see, for example,
\citealt{soucail01} and references therein; \citealt{bs01} and
references therein). Since the
effects of gravitational lensing depend upon the redshift of the source,
the cosmological background, and the distribution of matter in the universe,
they can be used to constrain the cosmological parameters and the 
primordial power spectrum of density fluctuations from which structure
originates. In addition, many of the effects produced by gravitational lenses,
such as image multiplicity, separations, and time delay,
depend strongly upon the matter distribution inside the lenses.
Hence, measurements of these effects can
provide a unique tool for probing the matter
distribution inside collapsed objects like galaxies and clusters,
providing the only direct measurement of their dark matter content, 
and constraining the theory of their 
formation and evolution.

Until recently, the internal structure of halos adopted in lensing
studies was generally some gravitational equilibrium 
distribution, either singular or
nonsingular (e.g., King model, singular isothermal sphere, 
pseudo-isothermal sphere), 
not necessarily motivated directly by the
theory of cosmological halo formation 
(see, e.g., \citealt{young80,tog84,hk87,nw88,blandford91,
jaroszynski91,jaroszynski92,kochanek95,pmm98,premadi01a,premadi01b,rm01}). 
As the theory
of halo formation in the CDM model has advanced in recent years, however,
the halo mass profiles adopted for lensing models have been refined to
reflect this theory.
Numerical simulations of large-scale
structure formation in Cold Dark Matter (CDM) universes predict that
galaxies and clusters have a singular density profile
which approaches a power law $\rho\propto r^{-n}$ at the center,
with the exponent $n$ ranging from 1 to 1.5 
\citep{cl96,nfw96,nfw97,tbw97,fm97,fm01a,fm01b,
moore98,moore99,hjs99,ghigna00,js00,klypin00,power02}.
These results are in apparent conflict with observations
of rotation curves of dark-matter-dominated dwarf 
galaxies and low surface brightness galaxies,
which favor a flat-density core (cf. \citealt{primack99,bs99,moore99,moore01}).
On the scale of clusters of galaxies, observations of 
strong gravitational lensing
of background galaxies by foreground clusters also favor the presence of a
finite-density core in the centers of clusters (see, e.g.,
\citealt{tkd98}). 

Several possible explanations have been suggested in order
to explain this discrepancy. 
The rotation curve data might lack sufficient spatial resolution 
near the center to distinguish unambiguously between a density profile 
with a flat-density core and one with a singular profile
(e.g. \citealt{vs01}).
Attempts have also been made to improve the numerical resolving 
power of the simulations to obtain a more accurate determination
of the slope of the predicted density profiles at small radii
(e.g. \citealt{moore99,power02}).
However, if the flat-core interpretation of the
observations and the singular cusps predicted by the numerical simulations are
both correct, then the simulation algorithms may be ignoring
some physical process which would, if included, serve to
flatten the halo density profiles at small 
radii relative to the results for purely gravitational, N-body
dynamics of cold, collisionless dark matter,
while retaining the more successful aspects of the
CDM model. For example, gasdynamical processes which involve dissipation
and radiative cooling, and perhaps energy-release feedback associated
with stars and quasars
(see, e.g. \citealt{gs99,esh01}) or modifications
of the microscopic properties of the dark matter, such as 
self-interacting dark matter
\citep{ss00,yoshida00,hs00,dave01}, warm dark matter \citep{cav00,sd01},
fluid dark matter \citep{peebles00}, fuzzy dark matter
\citep{hbg00}, decaying dark matter \citep{cen01}, 
annihilating dark matter \citep{kkt00}, or
repulsive dark matter \citep{goodman00}, all
have the potential to lower the central
density of halos and possibly reconcile simulations with observations.

Lensing by the two kinds of halo mass profiles, singular versus flat-core,
will be different. This has led to attempts to predict the
differences expected if the halos have the
singular cusp of the Navarro-Frenk-White (NFW) or Moore profiles or else
a profile with a flat core
(e.g. \citealt{kochanek95,km01,rm01,wts01,tc01,lo02}). 
Singular profiles like that of NFW are physically motivated by the N-body
simulations, and the latter have been used to place these halo
profiles empirically in a proper
cosmological context which permits statistical predictions for the
CDM model. The nonsingular profiles which have been adopted to contrast
with these singular ones, however,
are generally no more than parameterized, mathematical fitting formulae, with
no particular physical model to motivate them or put them in a proper
cosmological context.

We have developed an analytical model for the postcollapse equilibrium
structure of virialized objects that condense out of a cosmological background
universe, either matter-dominated or flat with a cosmological constant
(\citealt{sir99}, hereafter Paper~I;
\citealt{is01a}, hereafter Paper~II). This
{\it Truncated Isothermal Sphere\/}, or TIS, model assumes
that cosmological halos form from the collapse and virialization of
``top-hat'' density perturbations and are spherical, isotropic, and 
isothermal. This leads to a unique, nonsingular
 TIS, a particular solution of the Lane-Emden equation
(suitably modified when $\Lambda\neq0$). 
The size $r_t$ and velocity dispersion $\sigma_V$ are unique functions of the
mass $M$ and formation redshift $z_{\rm coll}$ of the object for
a given background universe. The TIS density profile flattens to a 
constant central value, $\rho_0$, which is roughly proportional to the
critical density of the universe at the epoch of collapse,
with a small core radius $r_0\approx r_t/30$
(where $\sigma_V^2=4\pi G\rho_0r_0^2$ and $r_0\equiv r_{\rm King}/3$,
for the ``King radius'' $r_{\rm King}$, defined by \citealt{bt87},
p. 228).

Even though the TIS model does not produce the central cusp in the
density profile of halos predicted by numerical CDM simulations at
very small radii,
it does reproduces many of the average properties of these halos 
quite well, suggesting that it is a useful 
approximation for the halos which result from more realistic initial 
conditions (Papers I, II; \citealt{is01b} and
references therein). In particular,
the TIS mass profile agrees well with the fit by NFW to
N-body simulations (i.e. fractional
deviation of $\sim20\%$ or less) at all radii outside of a few TIS core radii
(i.e. outside a King radius or so).
It also predicts the internal structure of X-ray
clusters found by N-body and gasdynamical simulations of cluster 
formation in the CDM model. For example, 
the TIS model reproduces to great accuracy the
mass-temperature and radius-temperature virial relations and integrated 
mass profiles derived empirically from the simulations of cluster formation
\citep{emn96}.
The TIS model also successfully reproduces to high precision the mass-velocity 
dispersion relation for clusters in CDM simulations of
the Hubble volume by the Virgo Consortium \citep{evrard02}, including 
its dependence on redshift for different background cosmologies. The
TIS model also 
correctly predicts the average value of the virial ratio
in N-body simulations of halo formation in CDM.   

The TIS profile matches the observed mass profiles of 
dark-matter-dominated dwarf galaxies. 
The observed rotation curves of
dwarf galaxies are generally well fit by a density profile
with a finite density core suggested by \citet{burkert95}, given by
\begin{equation}
\rho(r)=\frac
{\rho_{0,B}}{(r/r_c+1)(r^2/r_c^2+1)}\,.
\end{equation}

\noindent
The TIS model gives a nearly perfect fit to this profile,
with best fit parameters
$\rho_{0,B}/\rho_{0,{\rm TIS}}=1.216$, $r_{c}/r_{0,{\rm TIS}}=3.134$,
correctly predicting the maximum 
rotation velocity $v_{\rm max}$ and the radius $r_{\rm max}$
at which it occurs. The TIS model can also
explain the mass profile with a flat density core 
measured by \citet{tkd98} for cluster CL 0024+1654
at $z=0.39$, using the strong gravitational lensing of background 
galaxies by the cluster to infer the cluster mass distribution
\citep{si00}.
The TIS model not only provides a good fit to the projected 
surface mass density distribution of this cluster within the arcs, but
also predicts the overall 
mass, and a cluster velocity dispersion in close agreement with the value 
$\sigma_V=1150$ km/s measured by \citet{dressler99}.

Therefore, the TIS model can be applied to clusters of galaxies
or dark-matter-dominated dwarf galaxies, for which baryonic processes
have not significantly modified the mass profile.
At the intermediate scale of
large galaxies, where the central density profiles may be baryon-dominated,
following radiative cooling by the baryon gas, the central profiles may
differ from that of the halo of dark matter and baryons which would otherwise
form in the absence of radiative losses by the baryonic gas.
The TIS model ignores such processes,
just as the empirical density profiles (NFW, Moore, $\ldots$)
based on gravity-only N-body simulations do. Hence,
these profiles might not be directly applicable to galaxy-scale objects
for which the mass profiles have been significantly affected by such
baryonic processes.

Several authors have studied the effect of lensing by halos with
a flat-density core \citep{jaroszynski91,jaroszynski92,kochanek95,
pmm98,premadi01a,premadi01b} or by NFW or Moore profiles
that have been generalized, so
that the inner slope of the density profile is arbitrary
\citep{km01,rm01,wts01,lo02}.
These particular density profiles are essentially mathematical
conveniences without physical motivation. There is no underlying
theoretical model in these cases that was used to predict the value of the
core radius or the departure of the inner slope of the density profile
from the value found by N-body simulations of CDM.
By contrast, the TIS model is based on a set of physical
assumptions concerning the origin, evolution, and
equilibrium structure of halos in CDM universes. Observations of gravitational
lenses have the potential to distinguish between the TIS profile
and singular ones like the NFW profile, as several observable properties of 
gravitational lenses will be strongly affected by the presence, or
absence of a central cusp in the density profile. 
One example of an
important observable that can distinguish between various density profiles
is the parity of the number of images. Lenses with nonsingular density
profiles, such as the TIS, obey the {\it odd number theorem}. The number of
images of a given source is always odd, unless the source is extended and
saddles a caustic (see \citealt{sef92}, hereafter SEF,
p.~172). Lenses with singular profiles, like the singular isothermal sphere,
the NFW profile, or the Moore profile, need not obey this theorem, even for
point sources. Most observed multiple-image gravitational lenses
have either 2 or 4 images, and this may argue against 
profiles with a central core \citep{rm01}.
There are, however, other possible explanations for the 
absence of a third or fifth
image. That image tends to be very close to the optical axis, 
and might be hidden behind the lens itself. Also, it 
is usually highly demagnified, and might be too faint to be seen. 

We can use the TIS solution to model observed gravitational
lenses individually. Alternatively, we can use the
observations collectively to constrain the distribution of halo properties
as characterized by the TIS solution. These properties, core radius, velocity 
dispersion, central density, and so on, depend upon the mass of 
the lensing halos and
the redshift at which they form. Observational constraints on
the statistical distribution of these properties will, in turn, impose 
constrains on the cosmological parameters and the 
primordial power spectrum of density fluctuations.

In this paper, we derive all the lensing properties of the TIS. We also
compare the TIS with three other density profiles: The Navarro-Frenk-White
(NFW) density profile, the Singular Isothermal Sphere (SIS), 
and the Schwarzschild Lens\footnote{A point mass}.
To compare the lensing properties of these various lens models, we
focus on one particular cosmological model, the currently favored
COBE-normalized $\Lambda$CDM model with $\Omega_0=0.3$, $\lambda_0=0.7$,
and $H_0=70\,\rm km\,s^{-1}Mpc^{-1}$ (this model is also cluster-normalized).

The remainder of this paper is organized as follows. 
In \S2, we describe the TIS density profile and the various
comparison profiles.
In \S3, we derive
the lens equation. In \S4, we compute the critical curves and caustics.
In \S5, we study the properties of multiple images: separation,
magnification, brightness ratios, and time delay. In \S6, we study the 
properties of weak lensing, focusing on the average shear. 
In \S7, we discuss the likelihood of actually observing cases of strong
lensing produced by TIS halos.
Summary and conclusion are presented in \S8.

\section{THE DENSITY PROFILES}

In order to compare the predictions for halo lensing for different
halo density profiles, we must relate the parameters which define one profile
to those which define another. For this purpose, we shall assume that all halo
profiles contain the same total mass within a sphere of the same radius.
The TIS halo is uniquely specified by the central density $\rho_0$ and
core radius $r_0$. The TIS halo has a well-defined
outer radius $r_t$ at which the mass distribution is truncated, enclosing
a total mass $M_t$. There is a unique dimensionless density profile for the
TIS if radius and density are expressed in units of $r_0$ and $\rho_0$,
respectively. Since the central density $\rho_0$ is proportional to
$\rho_c(z_{\rm coll})$, the critical density at the epoch of its
formation, $z_{\rm coll}$, it is also possible to specify the profile by the
two parameters, total mass and collapse redshift, ($M_t$, $z_{\rm coll}$),
which is equivalent to specifying the pair ($r_0$, $\rho_0$). It is
customary to define the total mass of other profiles used to model CDM
halos as $M_{200}$, the mass inside a sphere of radius $r_{200}$ with a
mean density which is 200 times $\rho_c(z)$ at some redshift $z$. For
the sake of direct comparison with the TIS profile, we will fix $M_{200}$
and $r_{200}$ for halos of different profiles, which amounts to fixing $M_t$
and $z_{\rm coll}$ for the TIS halo, since $M_t=1.167M_{200}=772.6\rho_0r_0^3$,
$\rho_0=1.8\times10^4\rho_c(z_{\rm coll})$, and $r_{200}=24.2r_0$ if
$z=z_{\rm coll}$.

For the NFW profile, there is a third parameter, the concentration parameter 
$c$, which must be specified in addition to the parameters $M_{200}$ and 
$r_{200}$ (or, equivalently, $M_{200}$ and $z$). The value of $c$ is not
completely independent of the other parameters since there is a statistical
expectation that $c$ is correlated with $M_{200}$ and $z$. However,
for any individual halo, $c$ is not known a priori.

In what follows, we will consider two possibilities for comparing our TIS lens
with other halo lens models. In the first case, the assumptions will be made
that the lens redshift $z_L=z_{\rm coll}$ for the TIS halo, and the
concentration parameter $c$ for the NFW halo of the same mass $M_{200}$
and $r_{200}$ at that redshift will be the typical value expected from
the statistical correlation of $c$ with halo mass and the redshift of
observation of the halo, $z_{\rm obs}=z_L$.
In that case, the halos are fully specified
by the values of $M_{200}$ and $z_L$. In the second case, we can relax the
assumption that $z_L=z_{\rm coll}$. This makes the TIS halo a two-parameter 
model specified by ($M_{200}$, $z_{\rm coll}$), and it is assumed that the
halo which formed at some $z_{\rm coll}\geq z_L$ did not evolve between
$z_{\rm coll}$ and $z_L$. For the comparison NFW halo, we take the
same $M_{200}$ and $r_{200}$ as the TIS halo, but allow the
concentration parameter $c$ to take any value, not necessarily the typical one
for a halo of that mass $M_{200}$ at redshift $z_L$.
In what follows, we focus
primarily on the first of these cases. This will illustrate the
relative expectation of the TIS and other profile lenses in a sense
which is conservative in regard to the possible strength of the strong lensing
effects predicted by the TIS halos relative to the others.
In our discussion, \S7, we will then comment further on the effect of
relaxing the assumption that $z_L=z_{\rm coll}$, as discussed above.

\subsection{The Radial Density Profiles}

The density profile of the TIS is obtained numerically by solving a
differential equation. However, it is
well-fitted by the following approximation:
\begin{equation}
\label{tisfit}
\rho(r)=\rho_0\left({A\over a^2+r^2/r_0^2}-{B\over b^2+r^2/r_0^2}\right)
\end{equation}

\noindent (Paper~I, II) where $\rho_0$ is a characteristic density, $r_0$ is a 
characteristic radius, and $A=21.38$, $B=19.81$, $a=3.01$, $b=3.82$.
We will compute the lensing properties of halos with this density profile,
and compare them with the properties derived for three comparison profiles.
The first one is the Navarro, Frenk, and White (NFW) density profile,
\begin{equation}
\label{nfwfit}
\rho(r)={\rho_{\rm NFW}^{\phantom2}\over(r/r_{\rm NFW}^{\phantom2})
(r/r_{\rm NFW}^{\phantom2}+1)^2}\,,
\end{equation}

\noindent
where $\rho_{\rm NFW}^{\phantom2}$
is a characteristic density and $r_{\rm NFW}^{\phantom2}$ is a 
characteristic radius. We will also consider the Singular Isothermal Sphere
(SIS) density profile,
\begin{equation}
\label{sisfit}
\rho(r)={\sigma_V^2\over2\pi Gr^2}\,,
\end{equation}

\noindent where $\sigma_V^{\phantom2}$ is the velocity dispersion and $G$ is
the gravitational constant. This model might not represent actual halos
very well, but it is a well-studied profile that has important
theoretical value. Finally, for completeness, we will also consider
the Schwarzschild lens,
\begin{equation}
\label{schfit}
\rho(r)=M_{\rm Sch}\delta^3(r)\,
\end{equation}

\noindent where $M_{\rm Sch}$ is the lens mass and $\delta^3$ is the
three-dimensional delta function.

To compare the lensing properties of these various density profiles, we
use essentially the same approach as \citet{wb00}. 
The virial radius
$r_{200}$ of a halo located at redshift $z$ is defined, as usual, 
as being the radius inside which the mean density is equal to 200 times
the critical density $\rho_c(z)\equiv3H^2(z)/8\pi G$ at that redshift
[where $H(z)$ is the Hubble parameter]. The mass $M_{200}$ inside
that radius is given by
\begin{equation}
\label{m200}
M_{200}={800\pi\rho_c(z)r_{200}^3\over3}\,.
\end{equation}

\noindent
When comparing the lensing properties of different density profiles, we 
will consider halos that are located at the same redshift and have the same
value of $r_{200}$. By definition, these halos will also have the same
value of $M_{200}$. By stretching the terminology,
we will refer to $M_{200}$ as ``the mass of the halo.'' This point needs
to be discussed. For the Schwarzschild lens, $M_{200}$ is indeed the mass of
the halo. The SIS density profile drops as $r^{-2}$ at large $r$, and the
mass therefore diverges unless we introduce a cutoff. The halo mass will
then be equal to $M_{200}$ only if the cutoff is chosen to be $r_{200}$.
The NFW density profile drops as $r^{-3}$, hence the total mass diverges
logarithmically, and this profile also needs a cutoff.
The TIS density profile drops asymptotically as $r^{-2}$,
but the TIS model includes a cutoff. This cutoff is located a radius
$r_t\approx1.2r_{200}$, and the mass inside the cutoff is 
$M_t=1.168M_{200}$.
In any case, a rigorous definition of the halo mass would require an
unambiguous determination of the boundary between the halo and the
background matter (such determination exists only for the TIS model),
as well as dealing with the fact that the assumption of spherical symmetry
that enters in these models most likely breaks down for real halos at large
enough radii. Treating $M_{200}$ as being the actual mass of the halo
is the best compromise.

For a given halo mass $M_{200}$, redshift $z$, and cosmological background
model, all density profiles are fully determined. 
The parameters of the various profiles ($\rho_0$, $r_0$,
$\rho_{\rm NFW}^{\phantom2}$, $r_{\rm NFW}^{\phantom2}$,
$\sigma_V^{\phantom2}$, and $M_{\rm Sch}$) are computed as follows.
First, we compute $r_{200}$ using equation~(\ref{m200}).
The characteristic radius of the TIS is then given by
\begin{equation}
\label{r0}
r_0=r_{200}/\zeta_{200}=r_{200}/24.2
\end{equation}

\noindent 
(Paper~I). The value $\zeta_{200}=24.2$ is actually valid only
for the Einstein-de~Sitter model, but the dependence upon the
cosmological model is rather weak (Paper~II), and for simplicity
we shall use the same value for all models.
We now integrate equation~(\ref{tisfit}) between $r=0$ and $r=r_{200}$, and get
\begin{equation}
\label{m200tis}
M_{200}={4\pi\rho_0r_{200}^3K_{200}\over\zeta_{200}^2}\,,
\end{equation}

\noindent where

\begin{eqnarray}
\label{k200}
K_{200}\!\!\!\!&\equiv&\!\!\!\!
A-B-{aA\over\zeta_{200}}\arctan{\zeta_{200}\over a}
+{bB\over\zeta_{200}}\arctan{\zeta_{200}\over b}\nonumber\\
&&=2.144
\end{eqnarray}

\noindent 
(see also \citealt{ct02}, eq.~[10]). 
By combining equations~(\ref{m200}) and~(\ref{m200tis}), 
we can eliminate $M_{200}$ and $r_{200}$, and solve for $\rho_0$. We get
\begin{equation}
\label{rho0}
\rho_0={200\rho_c(z)\zeta_{200}^2\over3K_{200}}=1.8\times10^4\rho_c(z)\,.
\end{equation}

\noindent The characteristic radius of the NFW profile is given by
\begin{equation}
\label{rnfw}
r_{\rm NFW}^{\phantom2}=r_{200}/c\,,
\end{equation}

\begin{figure}
\includegraphics[width=84mm]{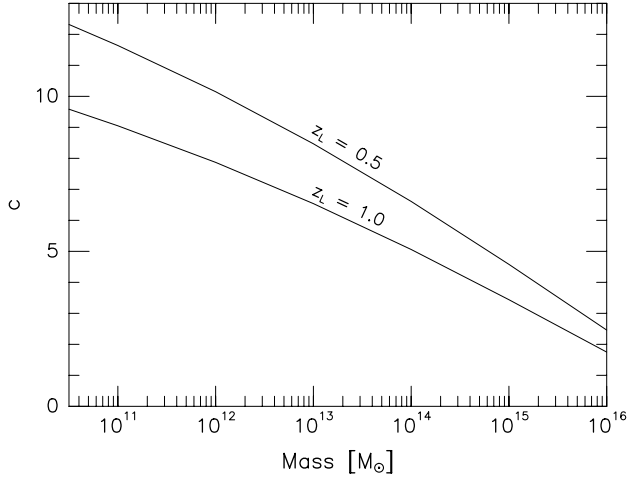}
 \caption{Concentration parameter $c$ versus halo mass, for NFW
halos located at redshifts $z_L=0.5$ and 1, as labeled.}
\label{ENS}
\end{figure}

\noindent where $c$ is the concentration parameter. Unlike the case for the
TIS, where $r_{200}/r_0$ is (nearly) constant,
the concentration parameter of the NFW profile is strongly dependent upon
the halo mass, redshift, and cosmological model. 
To determine $c$ for a given halo, 
we use the formalism of \citet{ens01}.
According to this formalism, the concentration parameter in the
$\Lambda$CDM model considered here varies with halo mass for two illustrative
values of halo lens redshift $z_L=0.5$ and 1.0
as shown in Figure~\ref{ENS}.
Once the value of $c$ is known, we can compute the parameters of the profile.
The characteristic density $\rho_{\rm NFW}^{\phantom2}$ is given by
\begin{equation}
\label{rhonfw}
\rho_{\rm NFW}^{\phantom2}={200c^3\rho_c(z)\over3[\ln(1+c)-c/(1+c)]}\,,
\end{equation}

\noindent
and the characteristic radius is given by equation~(\ref{rnfw}) above.
For the SIS, we integrate equation~(\ref{sisfit}) between $r=0$
and $r=r_{200}$, and get
\begin{equation}
\label{m200sis}
M_{200}={2\sigma_V^2r_{200}\over G}\,.
\end{equation}

\noindent Combining equations~(\ref{m200}) and (\ref{m200sis}), we get
\begin{equation}
\label{sigmav}
\sigma_V^{\phantom2}=\sqrt{400\pi G\rho_c(z)r_{200}^2\over3}\,.
\end{equation}

\begin{figure}
\vspace{-10pt}
\hspace{-15pt}
\includegraphics[width=94mm]{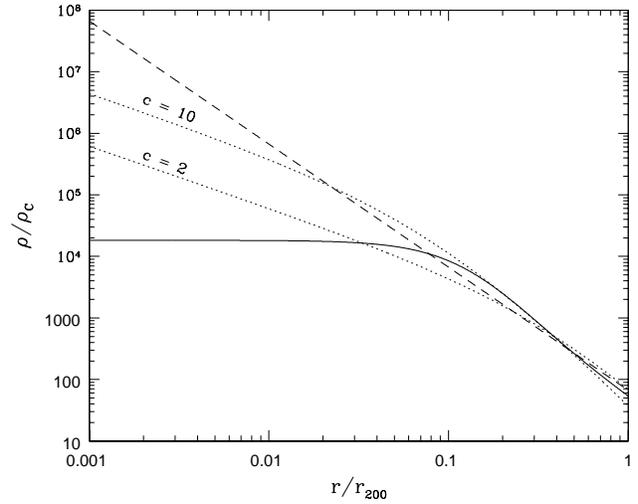}
 \vspace{-65pt}
 \caption{Radial density profiles, for 4 different halos with the same
values of $r_{200}$ and $M_{200}$. Solid curve: TIS; dotted curves:
NFW profiles with concentration parameters $c=2$ and 10 (as labeled);
dashed curve: SIS.}
\label{radprof}
\end{figure}

\noindent Finally, $M_{\rm Sch}$ is given by $M_{200}$. 
Figure~\ref{radprof} shows a comparison
of the various profiles, for halos with the same values of
$r_{200}$, and $M_{200}$.

\subsection{The Projected Surface Density}

The projected
surface density is given by
\begin{equation}
\label{proj}
\Sigma(\xi)=\int_{-\infty}^\infty\rho(r)dz\,,
\end{equation}

\noindent where $\xi$ is the projected distance from the center of the halo, 
and $z=(r^2-\xi^2)^{1/2}$. In principle,
this expression can be used only if our expressions for $\rho(r)$ 
(eqs.~[\ref{tisfit}]--[\ref{schfit}]) are valid all the way to $r=\infty$. 
As we pointed out in the previous section, most profiles must be truncated
at some finite radius $r_t$. For instance, the TIS is truncated at a
truncation radius $r_t\approx30r_0$ (the actual
value is 29.4 for an Einstein-de~Sitter universe [Paper I],
slightly different for an open matter-dominated universe, or a
flat universe with a cosmological constant [Paper~II]).
One could always set the limits in equation~(\ref{proj})
to $\pm(r_t^2-\xi^2)^{1/2}$. However, it turns out that
the resulting change in $\Sigma$ would be small, for all profiles considered.
To simplify the algebra, we shall assume that 
equation~(\ref{proj}) 
remains a good approximation out to $r=\infty$. We substitute 
equations~(\ref{tisfit})--(\ref{schfit}) in equation~(\ref{proj}). For
the TIS, we get
\begin{equation}
\label{sigma}
\Sigma_{\rm TIS}^{\phantom2}
(\xi)=\pi\rho_0r_0^2\left[{A\over\sqrt{a^2r_0^2+\xi^2}}-
{B\over\sqrt{b^2r_0^2+\xi^2}}\right]
\end{equation}

\noindent [This result was also derived by \citet{nl97},
and \citet{iliev00}].
For the NFW profile, we get
\begin{eqnarray}
\label{sigmanfw}
\Sigma_{\rm NFW}^{\phantom2}(\xi)\!\!\!\!&=&\!\!\!\!
2r_{\rm NFW}^{\phantom2}\rho_{\rm NFW}^{\phantom2}\nonumber\\
&&\kern-50pt\times\cases{
\displaystyle
{1\over x^2-1}\left[1-{2\over\sqrt{1-x^2}}
\arg\tanh\sqrt{1-x\over1+x}\right],&$\!\!\!x<1$;\cr
\noalign{\bigskip}
1/3,&$\!\!\!x=1$;\cr
\noalign{\bigskip}
\displaystyle
{1\over x^2-1}\left[1-{2\over\sqrt{x^2-1}}
\arctan\sqrt{x-1\over x+1}\right],&$\!\!\!x>1$;\cr}
\end{eqnarray}

\noindent where $x=\xi/r_{\rm NFW}^{\phantom2}$ \citep{bartelmann96,wb00}.
For the SIS, we get
\begin{equation}
\label{sigmasis}
\Sigma_{\rm SIS}^{\phantom2}(\xi)={\sigma_V^2\over2G\xi}\,.
\end{equation}

\noindent Finally, for the Schwarzschild Lens, we get
\begin{equation}
\label{sigmasch}
\Sigma_{\rm Sch}^{\phantom2}(\xi)=M_{\rm Sch}\delta^2(\xi)\,.
\end{equation}

\subsection{The Interior Mass Profile}

For spherically symmetric lenses, one important quantity is
the interior mass $M(\xi)$ inside a cylinder of projected radius
$\xi$ centered around the center of the lens. This quantity is given by
\begin{equation}
\label{minterior}
M(\xi)=2\pi\int_0^\xi\Sigma(\xi')\xi'd\xi'\,.
\end{equation}

\noindent
We have computed this expression for all density profiles considered,
using equations~(\ref{sigma})--(\ref{sigmasch}). We also reexpressed the
results in units of $M_{200}$ with the help of the equations given in
\S2.1. For the TIS, we get
\begin{eqnarray}
\label{minttis}
M_{\rm TIS}^{\phantom2}(\xi)\!\!\!\!&=&\!\!\!\!
{\pi M_{200}\over2\zeta_{200}K_{200}}
\big[A\sqrt{a^2+\xi^2/r_0^2}\nonumber\\
&&\qquad-B\sqrt{b^2+\xi^2/r_0^2}-Aa+Bb\big]\,.
\end{eqnarray}

\noindent
For the NFW profile, we get
\begin{equation}
\label{mintnfw}
M_{\rm NFW}^{\phantom2}(\xi)=4\pi r_{\rm NFW}^3\rho_{\rm NFW}^{\phantom2}
g(\xi/r_{\rm NFW}^{\phantom2})\,,
\end{equation}

\noindent where
\begin{equation}
g(x)=\cases{
\displaystyle
{2\over\sqrt{1-x^2}}\arg\tanh\sqrt{1-x\over1+x}+\ln{x\over2}\,,&$x<1$;\cr
\noalign{\bigskip}
\displaystyle
1+\ln{1\over2}\,,&$x=1$;\cr
\noalign{\bigskip}
\displaystyle
{2\over\sqrt{x^2-1}}\arctan\sqrt{x-1\over x+1}+\ln{x\over2}\,,&$x>1$;\cr}
\end{equation}

\noindent (\citealt{wb00}; \citealt{ct02}).
For the SIS, we get
\begin{equation}
\label{mintsis}
M_{\rm SIS}^{\phantom2}(\xi)={\pi M_{200}\xi\over2r_{200}}\,.
\end{equation}

\noindent Finally, for the Schwarzschild lens, we get
\begin{equation}
\label{mintsch}
M_{\rm Sch}^{\phantom2}(\xi)=M_{200}\,.
\end{equation}

\begin{figure}
\vspace{-15pt}
\hspace{-15pt}
\includegraphics[width=94mm]{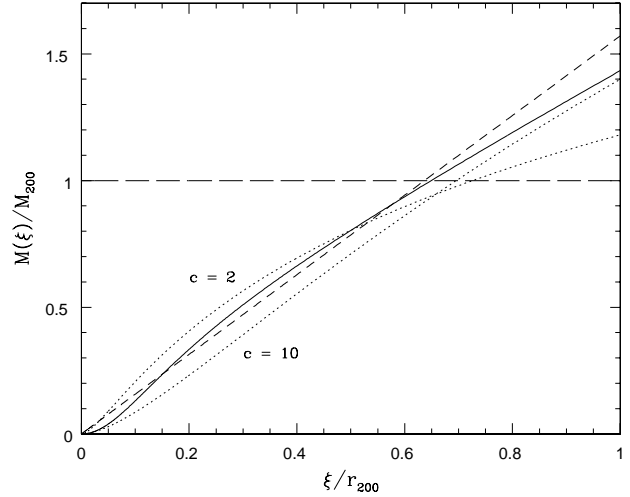}
 \vspace{-65pt}
 \caption{Interior mass profiles, for 5 different halos with the same
values of $r_{200}$ and $M_{200}$. Solid curve: TIS; dotted curves:
NFW profiles with concentration parameters $c=2$ and 10 (as labeled);
short-dashed curve: SIS; long-dashed curve: Schwarzschild lens.}
\label{massprof}
\end{figure}

\noindent Figure~\ref{massprof} 
shows a comparison of $M(\xi)$ for various halos with
the same values of $r_{200}$ and $M_{200}$.

\section{THE LENS EQUATION}

\begin{figure}
\vspace{-50pt}
\hspace{-45pt}
\includegraphics[width=106mm]{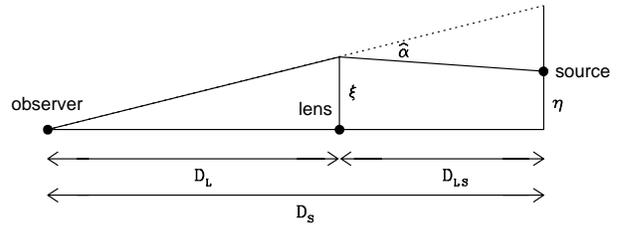}
 \vspace{-160pt}
 \caption{The lensing geometry: the dots indicate the location of the 
observer, lensing galaxy, and source. $\xi$
and $\eta$ are the positions of the image and the source, respectively, and
$\hat\alpha$ is the deflection angle.
The angular diameter distances $D_L$, $D_{LS}$, and $D_S$ are also indicated.}
\label{schematic}
\end{figure}

Figure~\ref{schematic} illustrates the lensing geometry. The quantities
$\eta$ and $\xi$ are the position
of the source on the source plane and the image on the image plane, respectively,
$\hat\alpha$ is the deflection angle, and $D_L$, $D_S$, and $D_{LS}$ are
the angular diameter distances between observer and lens, observer and source,
and lens and source, respectively. The lens equation is
\begin{equation}
\eta={D_S\over D_L}\xi-D_{LS}\hat\alpha
\end{equation}

\noindent 
[SEF, eq.~(2.15b)].
Notice that since the lens is axially symmetric, we can write the
quantities $\eta$, $\xi$, and $\hat\alpha$ as scalars instead
of 2-component vectors.
We introduce a characteristic length scale $\xi_0$, and nondimensionalize
the positions and deflection angle, as follows:
\begin{eqnarray}
y&=&{D_L\eta\over D_S\xi_0}\,,\\
\label{xscale}
x&=&{\xi\over\xi_0}\,,\\
\alpha&=&{D_LD_{LS}\hat\alpha\over D_S\xi_0}\,.
\end{eqnarray}

\noindent The lens equation reduces to
\begin{equation}
\label{lensfinal}
y=x-\alpha(x)\,.
\end{equation}

\noindent
For spherically symmetric lenses, the deflection angle is given by
\begin{equation}
\label{alpha1}
\alpha(x)={2\over x}\int_0^xx'{\Sigma(x')\over\Sigma_{\rm crit}}dx'
={2\over x}\int_0^xx'\kappa(x')dx'
\end{equation}

\noindent [SEF, eq.~(8.3)], where $\kappa\equiv\Sigma/\Sigma_{\rm crit}$
is the convergence, and
$\Sigma_{\rm crit}$ is
the critical surface density, given by
\begin{equation}
\label{sigmacrit}
\Sigma_{\rm crit}={c^2D_S\over4\pi GD_LD_{LS}}\,,
\end{equation}

\noindent where $c$ is the speed of light and $D_L$, $D_S$, and $D_{LS}$ are
the angular diameter distances between observer and lens,
observer and source, and lens and source, respectively. 
From equations~(\ref{minterior}) and (\ref{alpha1}), we get
\begin{equation}
\label{alpha1b}
\alpha(x)={M(\xi_0x)\over\pi\xi_0^2\Sigma_{\rm crit}x}\,.
\end{equation}

\noindent
Hence, the deflection angle is directly related to the
interior mass. For the TIS, we substitute
equation~(\ref{minttis}) into equation~(\ref{alpha1b}),
and set the 
characteristic scale $\xi_0$ equal to $r_0$. We get
\begin{eqnarray}
\label{alpha2}
\alpha_{\rm TIS}^{\phantom2}(x)\!\!\!\!&=&\!\!\!\!
{2\pi\rho_0r_0\over\Sigma_{\rm crit}x}
\big[A\sqrt{a^2+x^2}\nonumber\\
&&\qquad-B\sqrt{b^2+x^2}-Aa+Bb\big]\,.
\end{eqnarray}

\noindent This result was also obtained by \citet{ct01}.
We now introduce the
dimensionless central surface density, or central convergence, 
$\kappa_c$, defined by
\begin{equation}
\label{kappac}
\kappa_c\equiv{\Sigma(\xi=0)\over\Sigma_{\rm crit}}
={\pi\rho_0r_0\over\Sigma_{\rm crit}}\left({A\over a}-{B\over b}\right)\,,
\end{equation}

\noindent and use this expression to eliminate $\Sigma_{\rm crit}$ in
equation~(\ref{alpha2}). It reduces to
\begin{eqnarray}
\label{alpha3}
\alpha_{\rm TIS}^{\phantom2}(x)\!\!\!\!&=&\!\!\!\!
{2ab\kappa_c\over(Ab-Ba)x}
\big[A\sqrt{a^2+x^2}\nonumber\\
&&\qquad-B\sqrt{b^2+x^2}-Aa+Bb\big]\,.
\end{eqnarray}

\noindent For the NFW profile, we substitute equation~(\ref{mintnfw}) into
equation~(\ref{alpha1b}), and set 
$\xi_0=r_{\rm NFW}^{\phantom2}$. We get
\begin{equation}
\label{alphanfw}
\alpha_{\rm NFW}^{\phantom2}(x)={4\kappa_sg(x)\over x}
\end{equation}

\noindent where 
\begin{equation}
\label{kappas}
\kappa_s\equiv{\rho_{\rm NFW}^{\phantom2}
r_{\rm NFW}^{\phantom2}\over\Sigma_{\rm crit}}\,.
\end{equation}

\noindent
For the SIS profile, we substitute equation~(\ref{mintsis}) into
equation~(\ref{alpha1b}), and set
\begin{equation}
\label{xiosis} 
\xi_0={4\pi\sigma_V^2D_LD_{LS}\over c^2D_S}={\sigma_V^2\over G
\Sigma_{\rm crit}}\,.
\end{equation}

\noindent
We get
\begin{equation}
\alpha_{\rm SIS}^{\phantom2}(x)={x\over|x|}
\end{equation}

\noindent (SEF, \S8.1.4). Finally, for the
Schwarzschild lens, we set $M(\xi_0x)=M_{\rm Sch}$ and
\begin{equation}
\label{xiosch}
\xi_0=\sqrt{4GM_{\rm Sch}D_LD_{LS}\over c^2D_S}
=\sqrt{M_{\rm Sch}\over\pi\Sigma_{\rm crit}}\,. 
\end{equation}

\noindent
We get
\begin{equation}
\label{alphasch}
\alpha_{\rm Sch}^{\phantom2}(x)={1\over x}
\end{equation}

\noindent (SEF, \S8.1.2).

\section{CRITICAL CURVES AND CAUSTICS}

\subsection{Solutions}

The determination of the critical curves is quite trivial for
axially symmetric lenses. Equation~(\ref{alpha1b}) can be
rewritten as
\begin{equation}
\label{mx}
m(x)\equiv\alpha(x)x
\end{equation}

\noindent [SEF, eq.~(8.3)], where
$m(x)=M(\xi_0x)/\pi\xi_0^2\Sigma_{\rm crit}$ is the dimensionless
interior mass. Tangential and radial critical curves are 
defined respectively by
\begin{equation}
\label{tancrit}
{m(x_t)\over x_t^2}\equiv{\alpha(x_t)\over x_t}=1\,,
\end{equation}
\begin{equation}
\label{radcrit}
\left[{d(m/x)\over dx}\right]_{x=x_r}\equiv
\left[{d\alpha\over dx}\right]_{x=x_r}=1\,.
\end{equation} 

\begin{figure}
\hspace{-15pt}
\includegraphics[width=96mm]{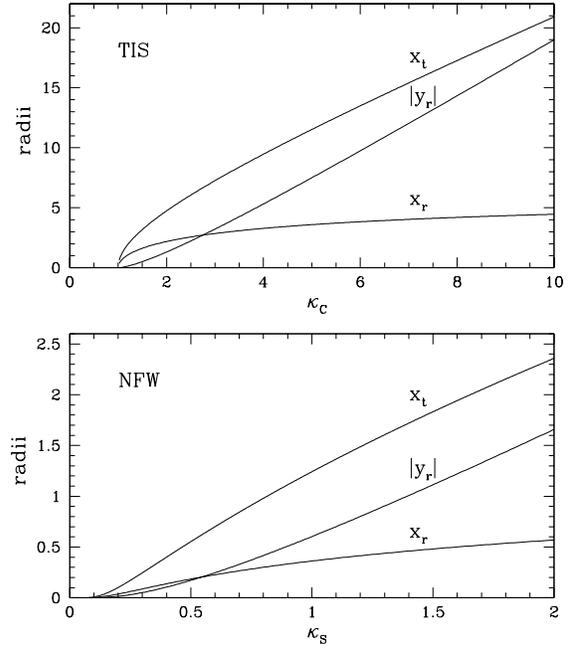}
 \vspace{-20pt}
 \caption{Top panel: Radii of the radial critical circle, $x_r$, tangential
critical circle, $x_t$, and radial caustic, $y_r$, versus $\kappa_c$,
for the TIS.
Bottom panel: $x_t$, $x_r$, and $y_r$ versus $\kappa_s$, for the NFW profile.}
\label{critfig1}
\end{figure}

\begin{figure*}
\vspace{-85pt}
\hspace{-35pt}
\includegraphics[width=156mm]{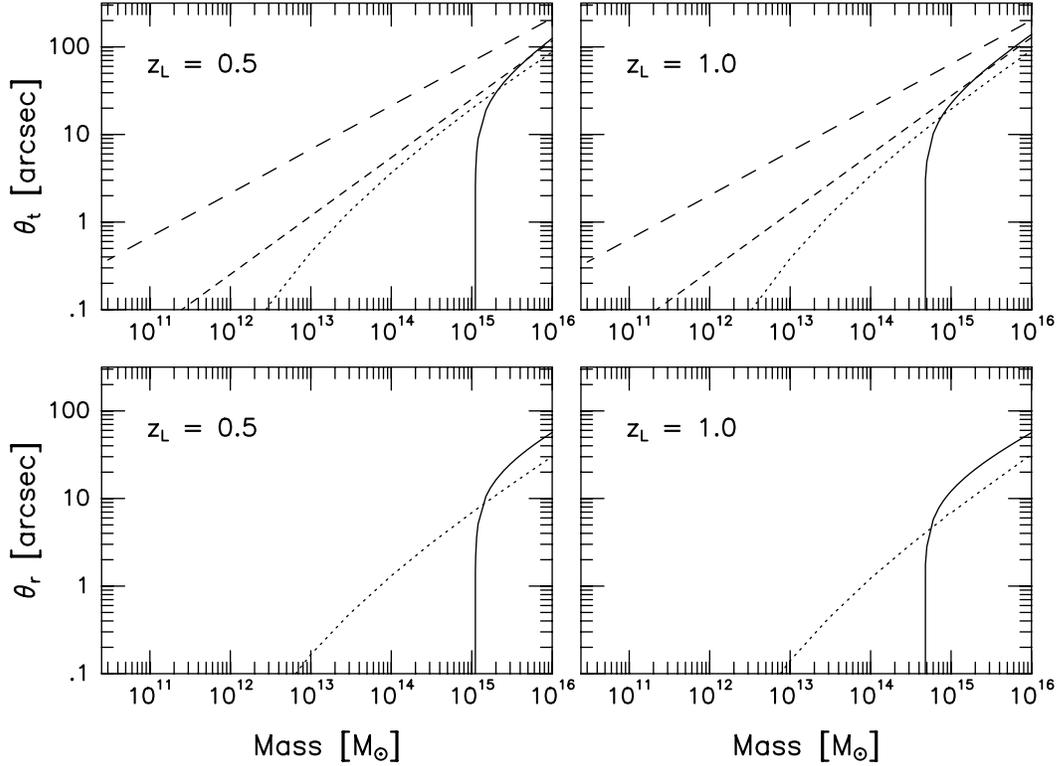}
 \vspace{-170pt}
 \caption{Angular radius of the tangential critical curve (top) and
radial critical curve (bottom), versus lens mass, for the
TIS (solid curves), the NFW profile (dotted curves), the SIS 
(short dashed curves) and the Schwarzschild lens (long dashed curves).
Results are shown for lenses located at redshift $z_L=0.5$ (left panels) and
1 (right panels).}
\label{critfig2}
\end{figure*}

For the Schwarzschild lens and the SIS, the solutions are $x_t=1$, and there
are no real solutions for $x_r$. For the NFW and TIS models, 
we have solved equations~(\ref{tancrit}) and~(\ref{radcrit}) numerically 
for $x_t$ and $x_r$. The solutions are plotted
in Figure~\ref{critfig1}, as functions of $\kappa_c$ and $\kappa_s$. 
Also plotted is the radial caustic
radius $y_r$, obtained by substituting the value of $x_r$ into 
equation~(\ref{lensfinal}).
(The value of $y_r$ we obtain is actually negative, because the
source and image are on opposite sides of the lens.
The actual radius of the caustic circle,
then, is the absolute value of $y_r$.)
Both $x_t$ and $|y_r|$
increase rapidly with $\kappa_c$ or $\kappa_s$, while the
value of $x_r$ levels off. Notice that there is no relation between $\kappa_c$
and $\kappa_s$. The former is the central convergence of the TIS, while the
latter is the convergence of the NFW profile at some finite radius
and its value depends upon the concentration parameter $c$.
Figure~\ref{critfig2} shows the angular radii of
the tangential and radial critical circles,
$\theta_r=\xi_r/D_L$ and $\theta_t=\xi_t/D_L$, in arc seconds as
functions of the mass of the lens, for lenses located at redshifts $z_L=0.5$
and 1. We assume that the source is located at redshift $z_S=3$. This is not a
a very constraining assumption, because the lensing properties vary weakly
with the source redshift for $z_S\gg1$. For instance, for lenses located
at redshifts $z_L\leq1$, the critical surface density $\Sigma_{\rm crit}$
varies by less than 20\% when the
source is moved from $z_S=3$ to $z_S=5$.

For spherically symmetric lenses, multiple images (and thus
critical circles) are possible only if the central convergence $\kappa(0)$
exceeds unity (SEF, p.~236, theorem~[e]). For the Schwarzschild lens, SIS,
and NFW profile, the central convergence diverges, hence these profiles
can always produce multiple images. But for the TIS, the central
convergence $\kappa(0)\equiv\kappa_c$ is finite, and multiple images
can be produced only if $\kappa_c>1$. This explains the sharp low-mass
cutoff seen in Figure~\ref{critfig2} for the TIS (solid curves).

The value of $\theta_t$ for the Schwarzschild lens is called the Einstein
radius $\theta_E$. It is often used to estimate the characteristic scale of
image features caused by strong lensing (e.g. ring radius, radial location
of arc, image separations) and to estimate the size of the region within
which the mass responsible for that strong lensing must be concentrated.
Since lensing halos are not actually point masses, however, the angular
radius $\theta_{\rm ring}$ of the 
actual Einstein ring which results if the source is located along the line 
of sight through the lens center will usually
differ from the Einstein radius $\theta_E$, assuming that the lens mass
distribution is actually capable of producing a ring. As we see in 
Figure~\ref{critfig2},
$\theta_E$ significantly exceeds $\theta_t$ for all profiles considered
(TIS, NFW profile, and SIS) for all masses considered. Hence, a mass estimate
based on assuming that the scale of image features is
of order $\theta_E$ will underestimate the actual mass of the lens, unless
the lens happens to be a Schwarzschild lens.

A source located behind the lens will produce multiple images if $y<y_r$.
The angular cross section for multiple imaging is therefore
\begin{equation}
\sigma_{\rm m.i.}=\pi\left({\eta_r\over D_S}\right)^2
=\pi\left({y_r\xi_0\over D_L}\right)^2\,.
\end{equation}

\begin{figure}
\vspace{-12pt}
\hspace{-15pt}
\includegraphics[width=90mm]{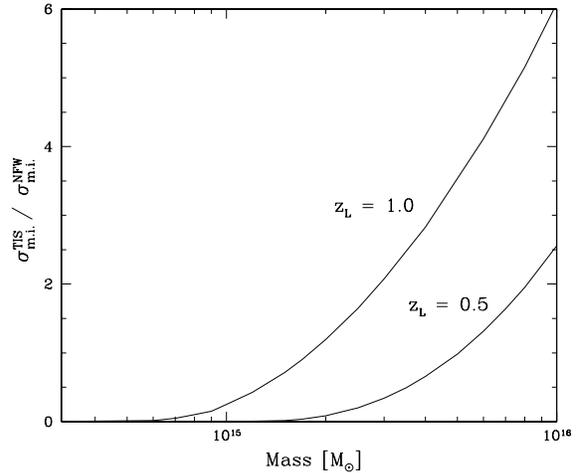}
 \vspace{-65pt}
 \caption{Ratio of the cross sections for multiple imaging by the TIS
and the NFW profile, versus lens mass, for lenses located 
at redshift $z_L=0.5$ and 1.}
\label{cross}
\end{figure}

\noindent In Figure~\ref{cross}, 
we plot the ratio of the cross sections for the
NFW and TIS profiles, for sources located at $z_S=3$. 
At low masses, $M<5\times10^{15}M_\odot$ for $z_L=0.5$,
$M<2\times10^{15}M_\odot$ for $z_L=1.0$, the ratios are less than unity, 
indicating that a distribution of lenses described by the NFW profile
would be more likely to produce cases with multiple images than if the same
distribution is described by the TIS model. This trend is reversed at
higher masses, and a TIS of mass $M=10^{16}M_\odot$ at $z_L=1.0$ is 
6 times more likely to produce multiple images than a NFW profile of
the same mass.

\begin{figure*}
\vspace{-30pt}
\hspace{-35pt}
\includegraphics[width=185mm]{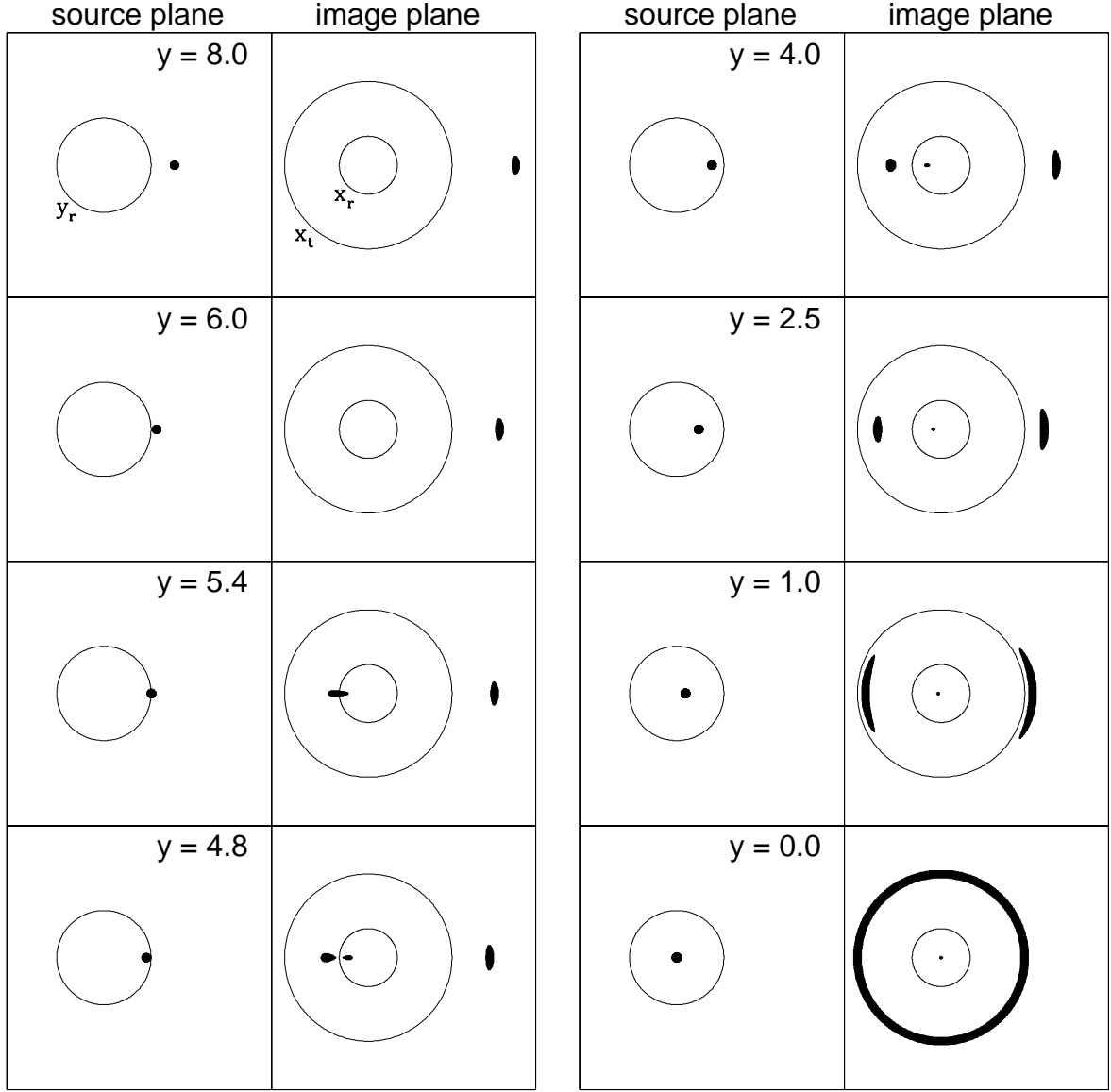}
\vspace{-55pt}
 \caption{Images of a circular source. Each pair of panels shows
the source plane in the left panel, with the caustic, and the
image plane in the right panel, with the radial (inner) and
tangential (outer) critical circles. The position $y$ of the source on
the source plane is indicated. We used $\kappa_c=4.014$, and a source
of diameter $\Delta y=1$.}
\label{images}
\end{figure*}

\subsection{Illustrative Example}

Using a simple ray-tracing algorithm, we computed the image(s) of a circular
source of diameter $\Delta y=1$, created by a TIS with central convergence
$\kappa_c=4.015$. The results are shown in Figure~\ref{images} for 8 different
locations of the source, ranging from $y=8.0$ to $y=0.0$. For each case, the
left panel shows the source and the caustic circle ($y_r=5.640$) on
the source plane, and the right panel shows the images(s), the radial
critical circle ($x_r=3.334$), and the tangential critical circle
($x_t=9.783$) on the image plane. For the cases $y=8.0$ and $6.0$,
only one image appears. At $y=5.4$, the source overlaps the caustic, and a
second, radially-oriented image appears on the radial critical circle.
At $y=4.8$, the source is entirely inside the caustic, and the second
image splits in two images, located on opposite sides of the radial
critical circle, forming with the original image a system of 3 aligned images.
As the source moves toward $y=0$, the central image moves toward $x=0$
and becomes significantly fainter, while the other images move toward the
tangential critical circle and become bright, elongated arcs. At $y=0$,
the two arcs have merged to form an Einstein ring located on top of
the tangential critical circle, while the central image, very faint,
is still visible in the center.

In the case of the TIS and the NFW profile, the radial caustics separate
regions on the source plane according to the number of images a source
in those regions produces, and when a source moves across such caustic, the
number of images changes. A curve on the
source plane that has this property is not always a caustic,
although in the case of the NFW and TIS profiles, it is.
A SIS will produce two images if $y<1$, and one image if $y>1$. However, the
circle defined by $y=1$ is not a caustic. When a source moves inside that
circle, a second image appears at $x=0$, but this value of $x$
does not satisfy equation~(\ref{radcrit}), and in particular the magnification
does not diverge at $x=0$, as in the case of a critical curve, but
vanishes instead. 
Indeed, the total magnification varies smoothly and monotonically as a source
moves across the $y=1$ circle.
For the Schwarzschild lens, there is no radial caustic, and this lens always
produces two images, no matter where the source is. This
happens because $\alpha(x)$ diverges as $x\rightarrow0$ (eq.~[\ref{alphasch}]),
so no matter how large the angular separation between source and
lens is, some rays will always be deflected toward the observer. However, for
large separations, the second image is demagnified by an enormous factor.

\section{PROPERTIES OF MULTIPLE IMAGES}

\subsection{Image Separation}

The locations of the images are computed by solving the lens 
equation~(\ref{lensfinal}). For the Schwarzschild lens, the solutions are
\begin{equation}
x_{1,2}={y\pm\sqrt{y^2+4}\over2}
\end{equation}

\noindent (SEF, eq.~[8.28]). Hence, there are always two images,
with separation
\begin{equation}
\label{sepsch}
(\Delta x)_{\rm Sch}=\sqrt{y^2+4}\,.
\end{equation}

\noindent For the SIS, the solutions are
\begin{equation}
x_{1,2}=y\pm1
\end{equation}

\noindent (SEF, eq.~[8.34c]). Again, there are always two images,
with separation
\begin{equation}
(\Delta x)_{\rm SIS}=2\,.
\end{equation}

\begin{figure}
\vspace{-30pt}
\hspace{0pt}
\includegraphics[width=84mm]{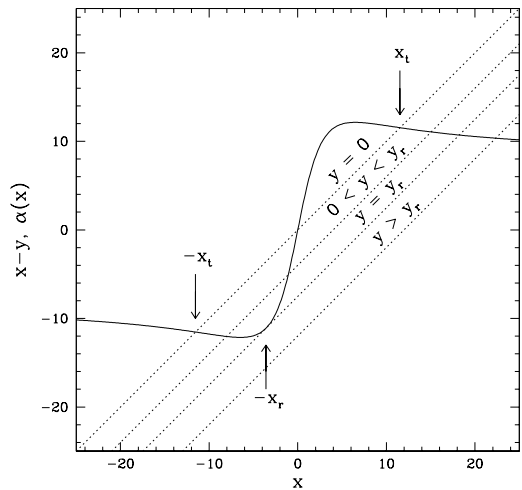}
 \vspace{-30pt}
 \caption{Plot of $\alpha(x)$ (solid curve) and $x-y$ (dotted lines)
versus $x$, for a TIS with $\kappa_c=5.005$ and 4 particular values
of $y$: $y=0$ (top line), $y=4$,
$y=y_r=7.515$, and $y=12$ (bottom line). 
Images are located at values of $x$ corresponding to
intersections between the lines and the curve. Particular solutions,
corresponding to images located on critical curves, are indicated by arrows.}
\label{multidiag}
\end{figure}

\noindent It is an interesting property of the SIS that the separation is
independent of the source location. For the NFW profile and the TIS, the
lens equation must be solve numerically.
In Figure~\ref{multidiag}, 
we show the {\it multiple image diagram} for the particular case
of a TIS with $\kappa_c=5.005$. The solid curves shows $\alpha(x)$, while the
dotted lines show $x-y$ for particular values of $y$. Each intersection
between a line and the curve corresponds to one image
(see equation~[\ref{lensfinal}]). If $y>y_r$ (bottom
line), the source is outside the caustic circle, and only one image appears.
If $y=y_r$ the source is on the caustic circle
and a second image appears on the radial
critical circle, at $x=-x_r$. For $y<y_r$, the source is inside the caustic,
and the second image splits into two images. Finally, for $y=0$ (top curve),
the central image is located at $x=0$, and the two outer images are located 
on the tangential circle, at $x=\pm x_t$. Actually, these two images merge
to form an Einstein ring. The slope of the curve $\alpha(x)$ versus $x$
is equal to $\kappa_c$ at $x=0$. It is clear from 
Figure~\ref{multidiag} that if we
lower $\kappa_c$ below 1, any $x-y$ versus $x$ line will intersect the
curve only once; multiple images cannot occur if $\kappa_c<1$. 
The corresponding diagram for the NFW profile is qualitatively similar.
However, in that case the slope of the curve $\alpha(x)$ versus $x$
diverges at $x=0$, hence there can always be 3 images, provided that 
the source is located inside the radial caustic

\begin{figure}
\vspace{-70pt}
\hspace{5pt}
\includegraphics[width=84mm]{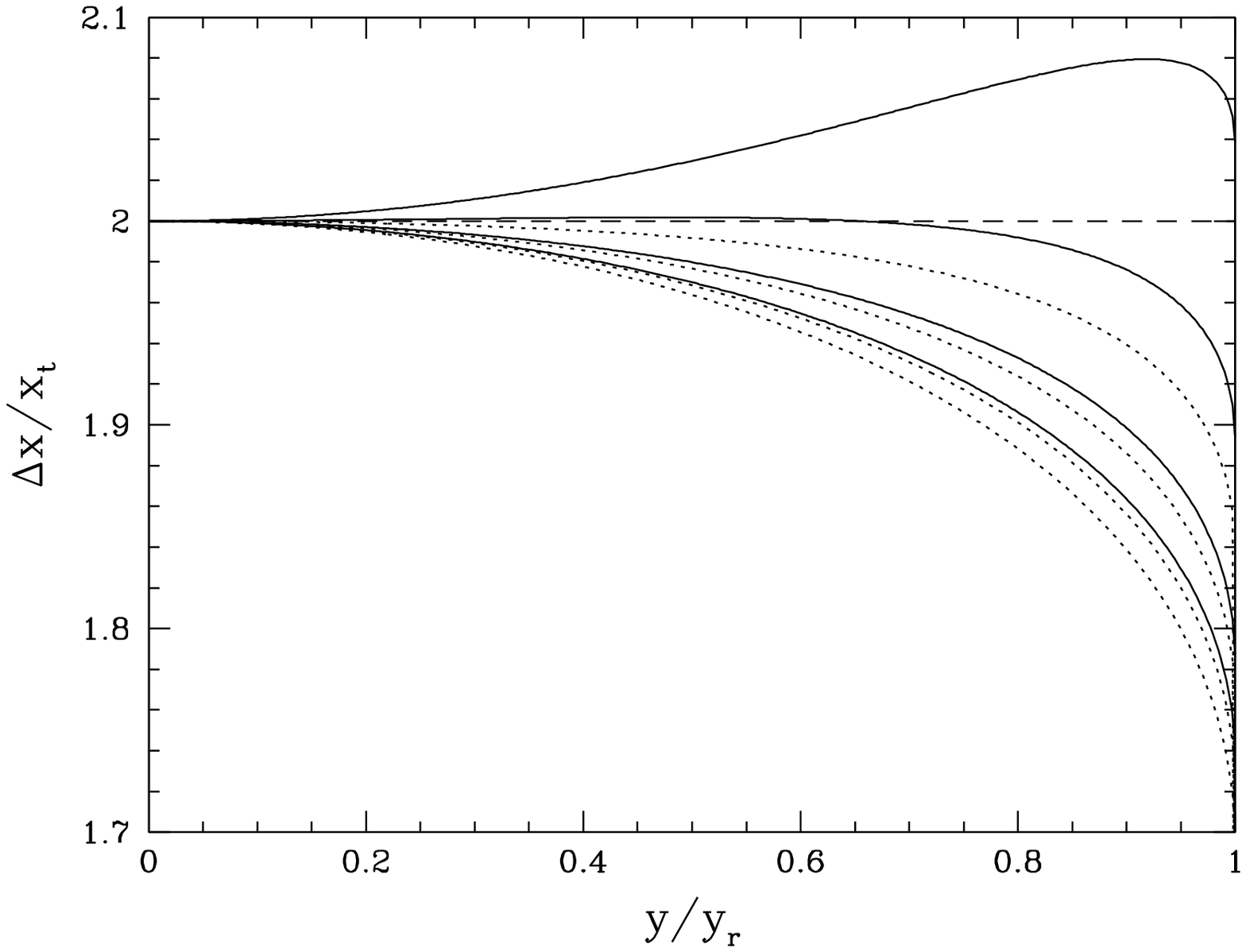}
 \vspace{-30pt}
 \caption{Separation $\Delta x$ between the two outer images,
in units of the tangential critical radius $x_t$, versus source
location $y$ in units of the caustic radius $y_r$. The solid curves, from top
to bottom, corresponds to TIS with $\kappa_c=10$, 5, 2.5, and 1.2, respectively.
The dotted curves, from top
to bottom, corresponds to NFW profiles 
with $\kappa_s=1.0$, 1.0, 0.5, and 0.2, respectively. 
The dashed line corresponds to the SIS. Results for the Schwarzschild lens are
not plotted.}
\label{separation}
\end{figure}

In Figure~\ref{separation}, 
we plot the separation between the two outer images as
a function of the source location. The plot only extends to $y/y_r=1$, since
larger values of $y$ only produce one image. 
The solid and dotted curves show the separations for the TIS and NFW profile,
respectively, with various values of $\kappa_c$ and $\kappa_s$.
The separation is fairly
insensitive to the source location, and stays within $\sim15\%$ of the
Einstein ring diameter $\Delta x=2x_t$ for all values of $\kappa_c$ and $\kappa_s$
considered. For the SIS and Schwarzschild lens, $y_r$ is undefined.
The dashed line in Figure~\ref{separation}
shows the constant separation for the SIS.
For the Schwarzschild lens
the separation depends upon $y$, and therefore cannot be plotted in
Figure~\ref{separation} since $y_r$ is undefined.
However, in the limit of large $y$, it can be shown that
the magnification of the faintest image drops as $1/y^4$ (see SEF,
eq.~[8.29a]). Hence, in practice, the second image will be visible
only if $y$ is small, in which case equation~(\ref{sepsch}) shows that
the separation is close to $2x_t$, since $x_t=1$ for that lens.

\begin{figure*}
\vspace{-40pt}
\hspace{-10pt}
\includegraphics[width=156mm]{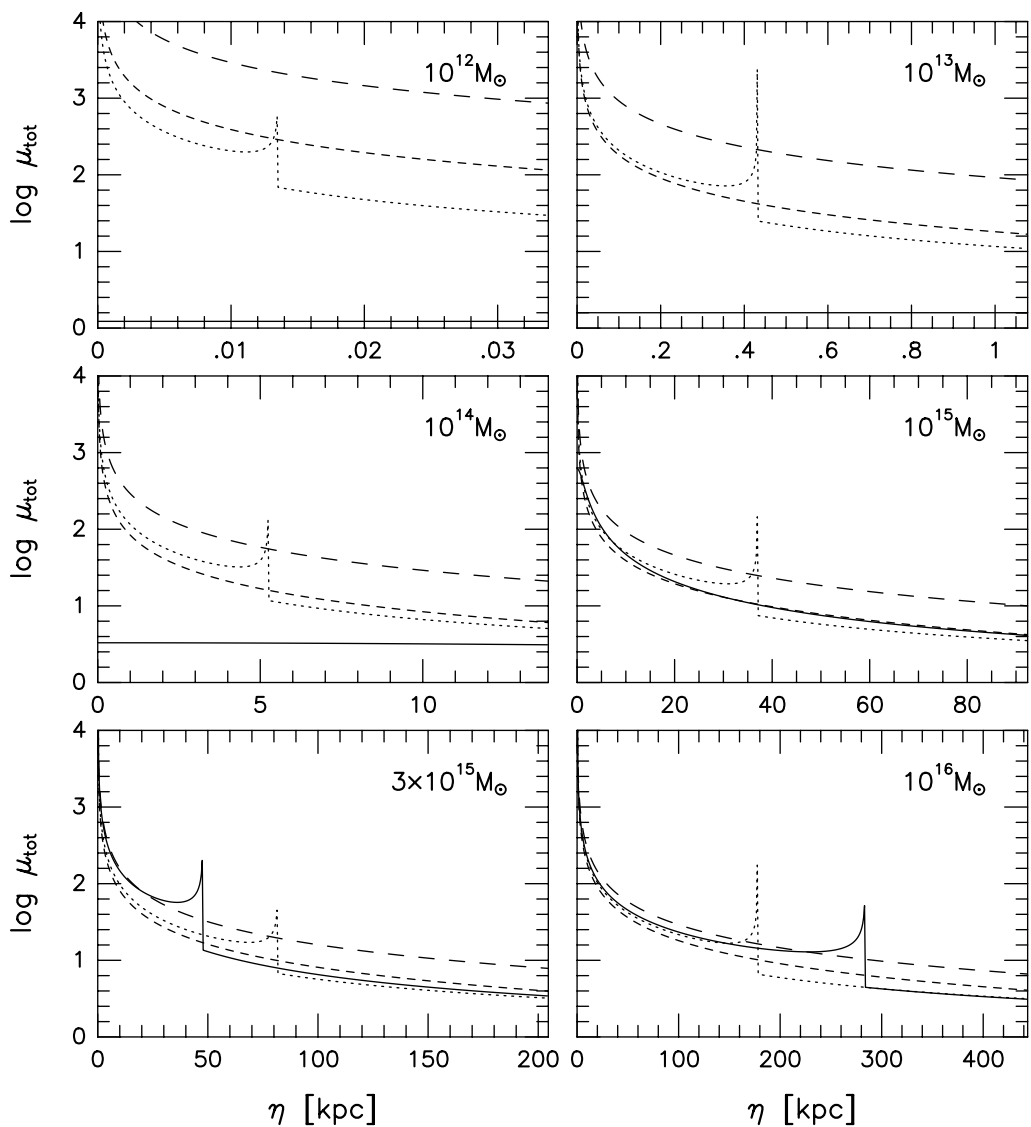}
 \vspace{-90pt}
 \caption{Total magnification $\mu_{\rm tot}$ versus source location $\eta$
for lenses of masses $10^{12}-10^{16}M_\odot$ located at $z_L=0.5$, for the
TIS (solid curves), the NFW profile (dotted curves), the SIS 
(short dashed curves) and the Schwarzschild lens (long dashed curves).}
\label{mag05}
\end{figure*}

Therefore, for all profiles considered, the image separation is
always of order the Einstein ring diameter, independently of the source location.
This is particularly convenient for theoretical studies, when
the actual source location can be ignored
(see, e.g., \citealt{mpm02}).

\subsection{Magnification and Brightness Ratios}

\begin{figure*}
\vspace{-40pt}
\hspace{-10pt}
\includegraphics[width=156mm]{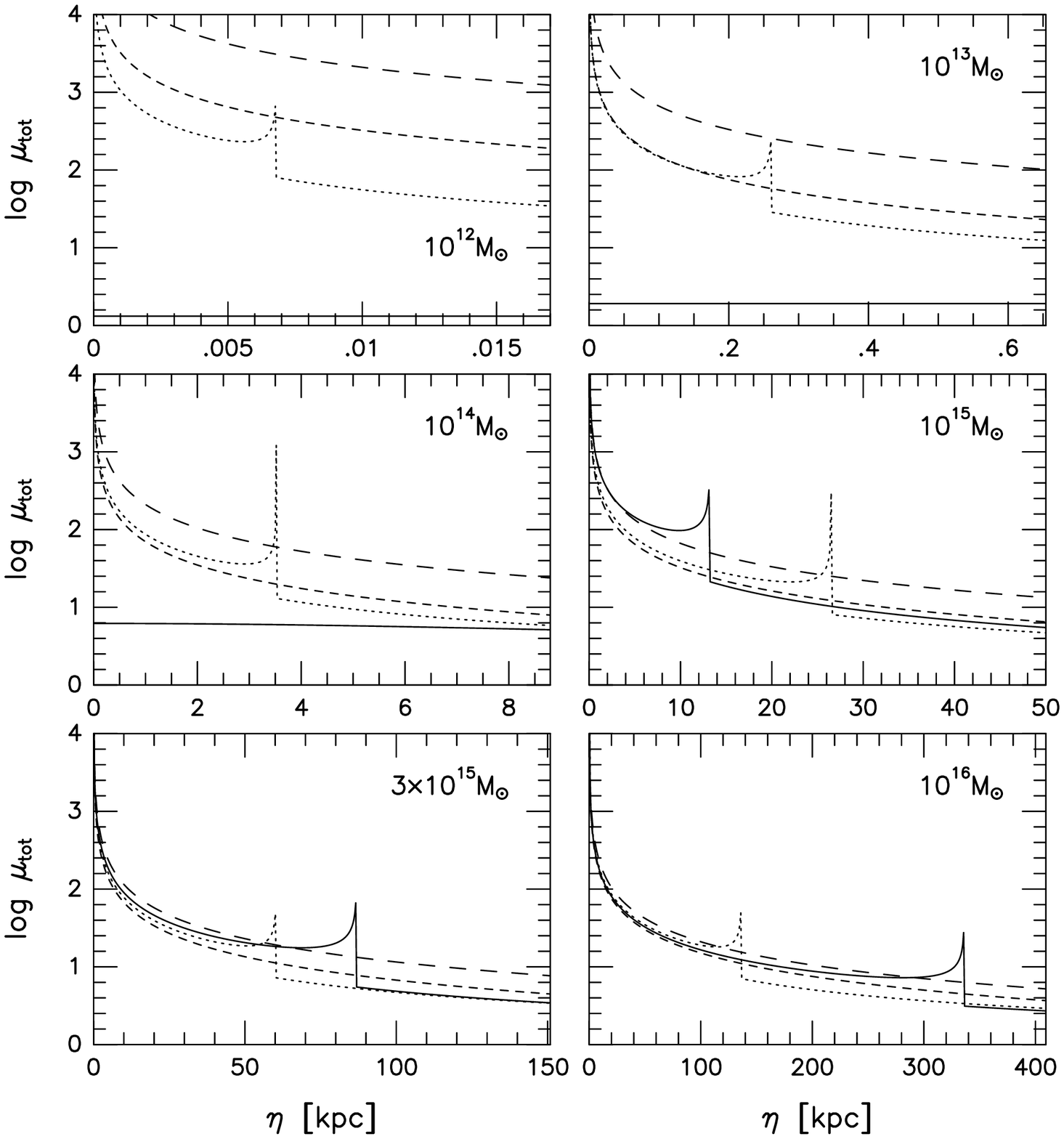}
 \vspace{-90pt}
 \caption{Total magnification $\mu_{\rm tot}$ versus source location $\eta$
for lenses of masses $10^{12}-10^{16}M_\odot$ located at $z_L=1.0$, for the
TIS (solid curves), the NFW profile (dotted curves), the SIS 
(short dashed curves) and the Schwarzschild lens (long dashed curves).}
\label{mag10}
\end{figure*}

\begin{figure*}
\vspace{-40pt}
\hspace{-10pt}
\includegraphics[width=156mm]{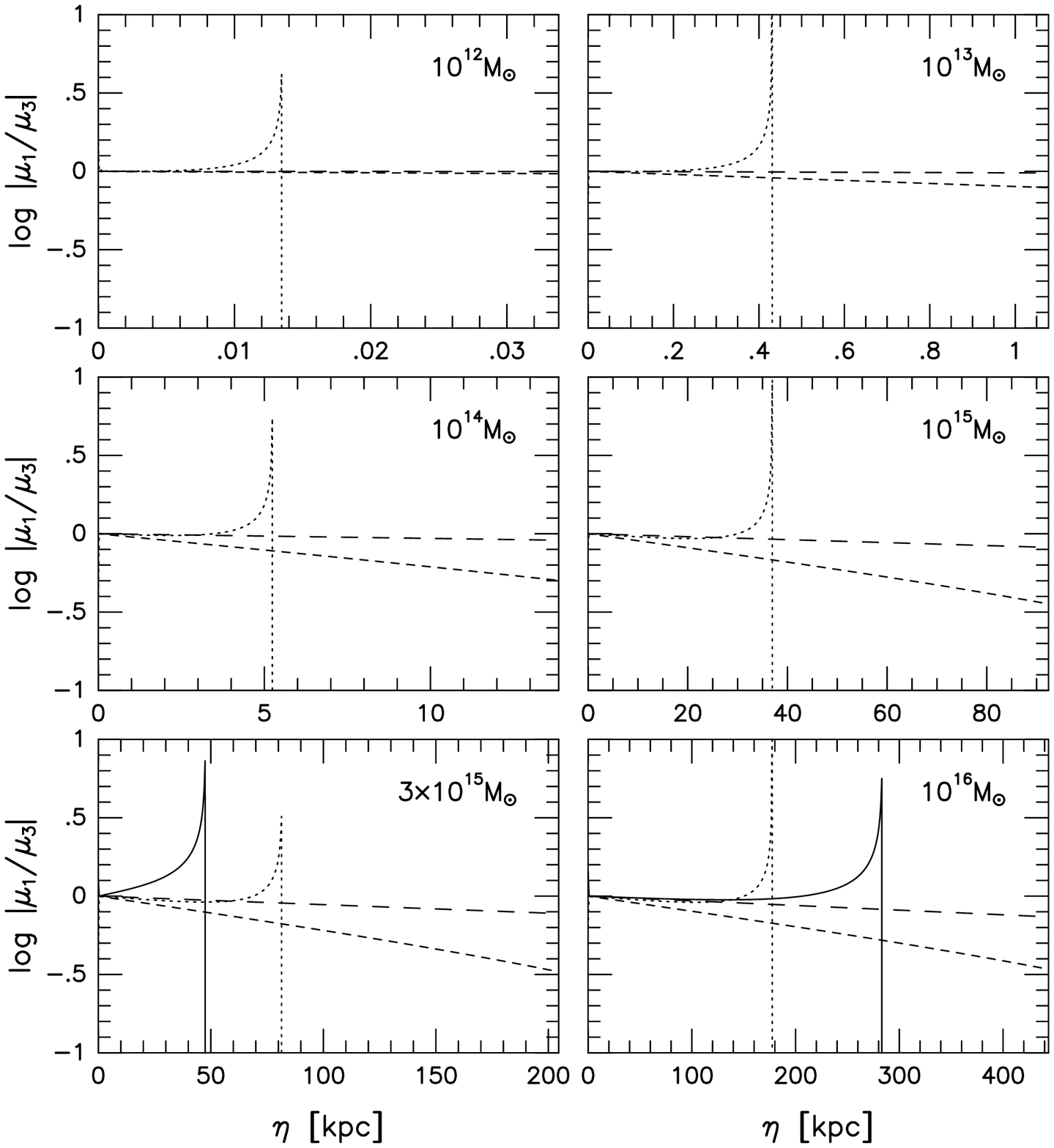}
 \vspace{-90pt}
 \caption{Brightness ratio $|\mu_1/\mu_3|$ versus source location $\eta$
for lenses of masses $10^{12}-10^{16}M_\odot$ located at $z_L=0.5$, for the
TIS (solid curves), the NFW profile (dotted curves), the SIS 
(short dashed curves) and the Schwarzschild lens (long dashed curves).}
\label{bratio05}
\end{figure*}

\begin{figure*}
\vspace{-40pt}
\hspace{-10pt}
\includegraphics[width=156mm]{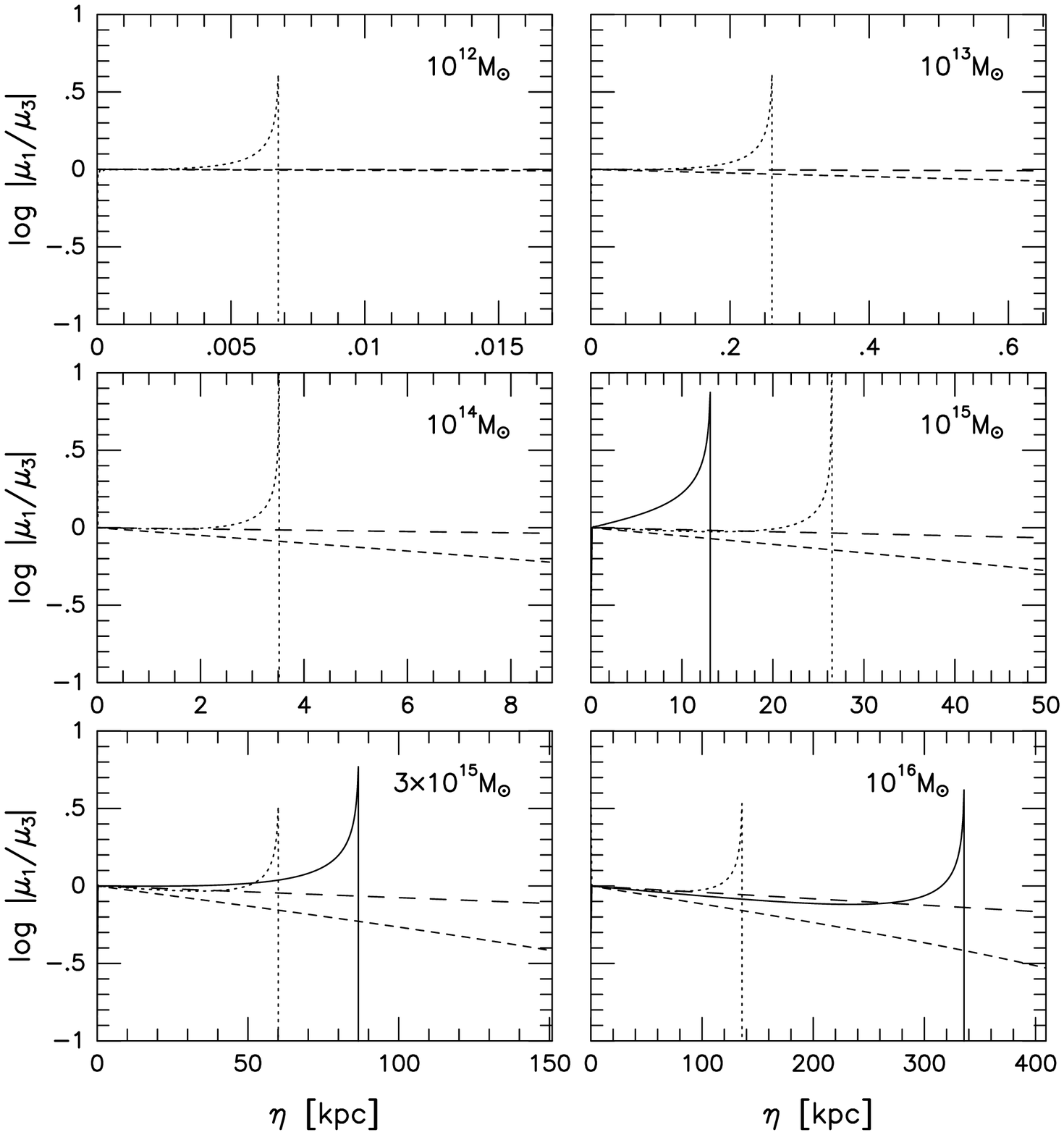}
 \vspace{-90pt}
 \caption{Brightness ratio $|\mu_1/\mu_3|$ versus source location $\eta$
for lenses of masses $10^{12}-10^{16}M_\odot$ located at $z_L=1.0$, for the
TIS (solid curves), the NFW profile (dotted curves), the SIS 
(short dashed curves) and the Schwarzschild lens (long dashed curves).}
\label{bratio10}
\end{figure*}

\begin{figure*}
\vspace{-40pt}
\hspace{-10pt}
\includegraphics[width=156mm]{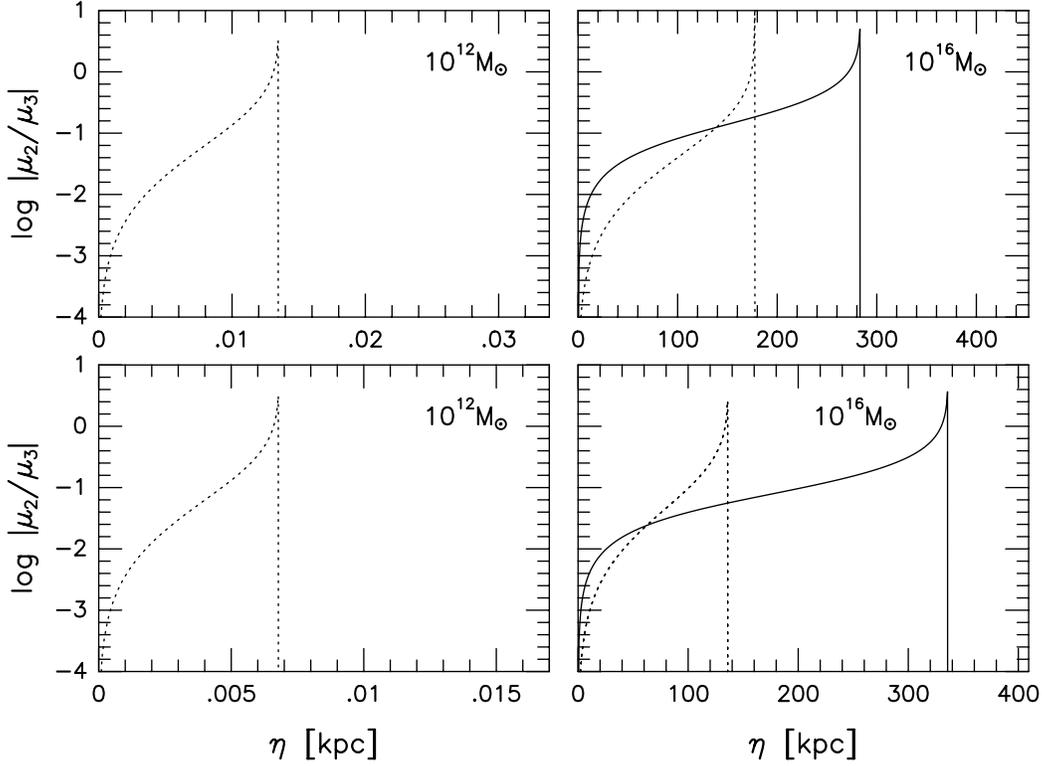}
 \vspace{-220pt}
 \caption{Brightness ratio $|\mu_2/\mu_3|$ versus source location $\eta$
for lenses of masses $10^{12}$ and $10^{16}M_\odot$ located at 
$z_L=0.5$ and 1, for the
TIS (solid curves) and the NFW profile (dotted curves).}
\label{bratio23}
\end{figure*}

The magnification of an image located at position $x$ on the lens
plane is given by
\begin{eqnarray}
\label{magni}
\mu\!\!\!\!&=&\!\!\!\!
\left(1-{m\over x^2}\right)^{-1}
\left[1-{d\over dx}\left({m\over x}\right)\right]^{-1}\nonumber\\
&&\qquad\qquad\qquad=\left(1-{\alpha\over x}\right)^{-1}
\left(1-{d\alpha\over dx}\right)^{-1}
\end{eqnarray}

\noindent (SEF, eq.~[8.17]).
We computed the magnification of the images produced by halos of masses
$10^{12}$, $10^{13}$, $10^{14}$, $10^{15}$, $3\times10^{15}$, 
and $10^{16}M_\odot$ located
at redshifts $z_L=0.5$ and $z_L=1$, for sources located
at redshift $z_S=3$. The results are plotted in 
Figures~\ref{mag05}--\ref{bratio23}
as functions of the source position. Figures~\ref{mag05} and \ref{mag10} show
the total magnification.
The dotted curves show the results for a NFW profile. As $y$ (or $\eta$)
decreases, the magnification
slowly increases, until the source reaches the radial caustic $y=y_r$.
At that moment, a second image, with infinite magnification 
($1-d(m/x)/dx=0$ in eq.~[\ref{magni}]) 
appears on the radial critical curve (for clarity, we truncated those infinite
``spikes'' in Figs.~\ref{mag05} and \ref{mag10}). As $y$
keeps decreasing, that second image splits into two images, and the 
total magnification becomes finite again, until the source reaches
$y=0$, and an Einstein ring with infinite magnification
($1-m/x^2=0$ in eq.~[\ref{magni}]) appears on the tangential critical curve.
Of course, these infinite magnifications are not physical, since they can only
occur for point sources.
The total magnification is always larger than unity, and always larger
when 3 images are present.

The solid curves in Figures~\ref{mag05} and \ref{mag10} 
show the results for the TIS.
At low masses, there is no radial caustic (See Fig.~\ref{critfig2}),
and only one image
appears. Because of the presence of a flat density core, the magnification
is nearly constant if the path of the rays goes near the center of the core.
For instance, for the case $M_{200}=10^{12}M_\odot$, $z_L=0.5$ (top left
panel of Fig.~\ref{mag05}), $r_0=7.125\,{\rm kpc}$, hence,
over the range of $\eta$ being plotted, we are way inside the core. As
the mass increases, the magnification increases, until a radial
caustic forms. This happens at $M_{200}=1.11\times10^{15}M_\odot$
for $z_L=0.5$, and at $M_{200}=4.7\times10^{15}M_\odot$ for $z_L=1$,
according to Figure~\ref{critfig2}.
At this point, the TIS has the ability to form
three images, and the results become qualitatively similar to the ones
for the NFW profile.

The short-dashed and long-dashed curves in Figures~\ref{mag05} and
\ref{mag10} show the
results fot the SIS and the Schwarzschild lens, respectively. Because of
the absence of radial caustic, the magnification always varies smoothly with
source position, and the only divergence occurs at $\eta=0$, where
an Einstein ring forms.

For a given mass, the Schwarzschild lens always produces the strongest
magnification, unless the source is very close to the radial caustic of a TIS
or a NFW profile, where a ``spike'' of infinite magnification
forms. At small masses ($M_{200}=10^{12}M_\odot$), the SIS
produces a stronger magnification than the NFW profile, for all source
positions. However, at larger masses, we find a different behavior:
If the NFW profile produces only one image (that is, we are on the right
hand side of the dotted spike in Fig.~\ref{mag05} and \ref{mag10}),
the SIS still
produces a larger magnification than the NFW profile. But if the NFW profile
produces 3 images, then the total magnification exceeds the one produced by
the SIS. As for the TIS, at low masses, where only one image forms, the
magnification is much smaller than for the other profiles. But at large
masses, where multiple images can form, the magnification becomes comparable
to the one for the other profiles.

Figures~\ref{bratio05}, \ref{bratio10}, and \ref{bratio23}
show the brightness ratios $|\mu_1/\mu_3|$,
and $|\mu_2/\mu_3|$, where $\mu_i$ is the magnification
of image~$i$, as a function of source position. By convention, image 1 is
the one between the tangential and radial critical curves, image 2
is the one inside the radial critical curve, and image 3 is the one
outside the tangential critical curve (see Fig.~\ref{images};
from left to right,
the three images are image 1, 2, and 3). The ratio $|\mu_1/\mu_3|$,
plotted in Figures~\ref{bratio05} and~\ref{bratio10}, always
goes to unity in the limit $\eta\rightarrow0$, as images 1 and 3 merge to
form an Einstein ring. For $\eta>0$, image 1 is usually fainter than image 3
($|\mu_1/\mu_3|<1$), unless it is located near the radial critical curve,
in which case $\mu_1$ diverges. For the ranges of $\eta$ being
considered, The ratio $|\mu_1/\mu_3|$ is always very
close to unity for the Schwarzschild lens, and fairly close to unity for the
SIS. The TIS and NFW profile do not produce an image 1 at large $\eta$,
where the source is outside of the radial caustic. Notice that in these
figures no curves are plotted for
the TIS at small mass, because only one image forms.

The ratio $|\mu_2/\mu_3|$, plotted in Figure~\ref{bratio23}
for a few cases, is much
less interesting. Central image 2, which can be produced only with the TIS or
NFW profile, is usually very faint unless it is located near the radial
critical curve, in which case it might be too close to image 1 to
be resolved individually.

\subsection{Shear}

\begin{figure*}
\vspace{-20pt}
\includegraphics[width=156mm]{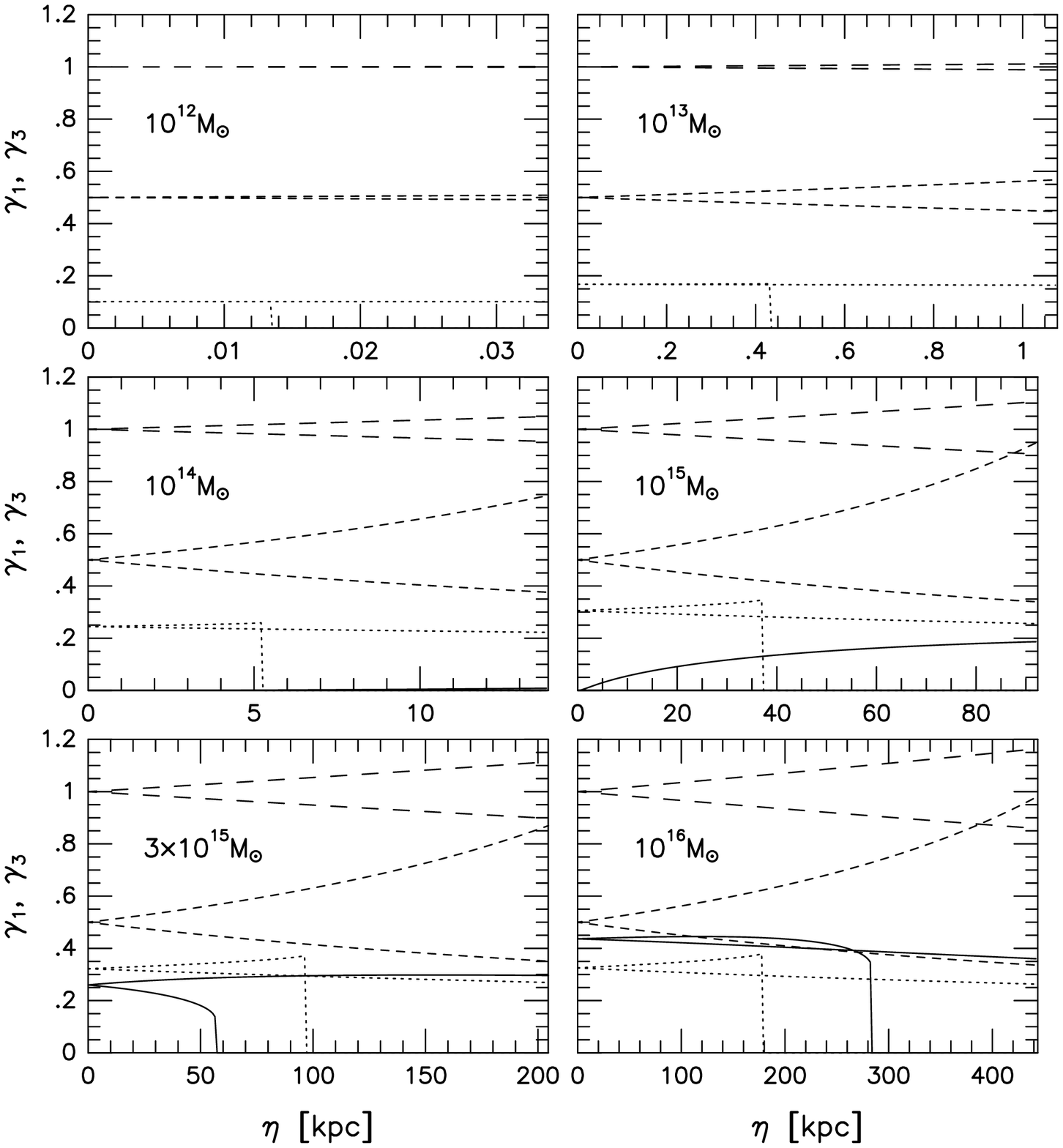}
 \vspace{-80pt}
 \caption{Shear $\gamma_1$ and $\gamma_3$ of images 1 and
3, versus source location $\eta$
for lenses of masses $10^{12}-10^{16}M_\odot$ located at $z_L=0.5$, for the
TIS (solid curves), the NFW profile (dotted curves), the SIS 
(short dashed curves) and the Schwarzschild lens (long dashed curves).
When two curves of the same type appear, the top one is $\gamma_1$,
the bottom one $\gamma_3$.}
\label{shear05}
\end{figure*}

\begin{figure*}
\vspace{-20pt}
\includegraphics[width=156mm]{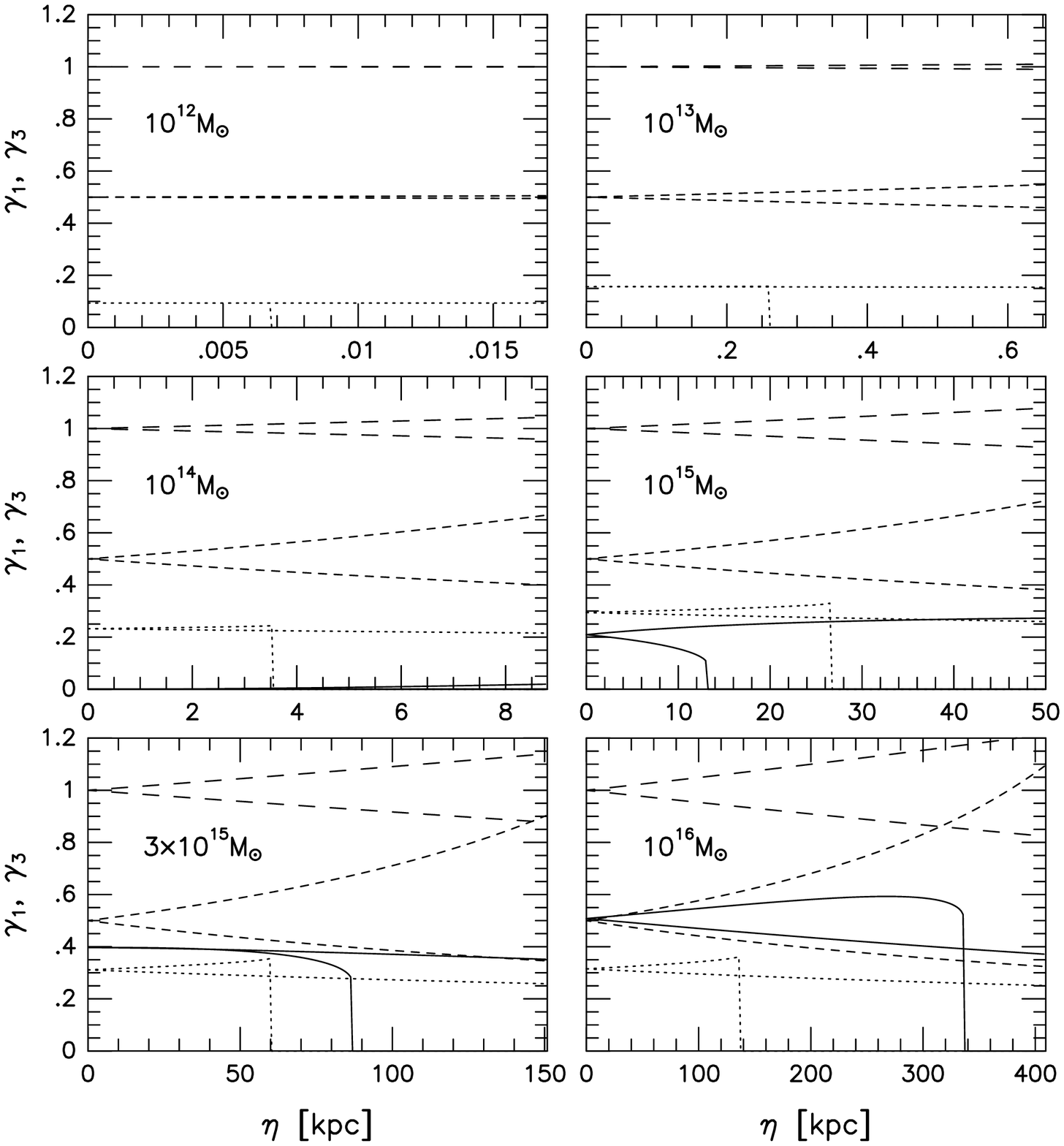}
 \vspace{-80pt}
 \caption{Shear $\gamma_1$ and $\gamma_3$ of images 1 and
3, versus source location $\eta$
for lenses of masses $10^{12}-10^{16}M_\odot$ located at $z_L=1.0$, for the
TIS (solid curves), the NFW profile (dotted curves), the SIS 
(short dashed curves) and the Schwarzschild lens (long dashed curves).
When two curves of the same type appear, the top one is $\gamma_1$,
the bottom one $\gamma_3$.}
\label{shear10}
\end{figure*}

The total shear $\gamma(x)$ of an image located at position $x$ is given by
\begin{equation}
\label{shear}
\gamma=\left|{m(x)\over x^2}-\kappa(x)\right|\,,
\end{equation}

\noindent (SEF, eq.~[8.15]). For the TIS, we substitute 
equations~(\ref{sigma}) and~(\ref{mx}) into equation~(\ref{shear}), and get
\begin{eqnarray}
\gamma_{\rm TIS}^{\phantom2}(x)\!\!\!\!&=&\!\!\!\!{ab\kappa_c\over Ab-Ba}\bigg|
{2A\over x^2}\left[\sqrt{a^2+x^2}-a\right]\nonumber\\
&&\kern-30pt-{2B\over x^2}\left[\sqrt{b^2+x^2}-b\right]
-{A\over\sqrt{a^2+x^2}}+{B\over\sqrt{b^2+x^2}}\bigg|\,.
\end{eqnarray}

\noindent [this result was also obtained by \citet{nl97}].
For the NFW profile, the shear is given by
\begin{equation}
\gamma_{\rm NFW}^{\phantom2}(x)=\kappa_s\times\cases{
g(x)\,,&$x\neq1$\,;\cr\noalign{\vspace{5pt}}
\displaystyle{10\over3}-4\ln2\,,&$x=1$\cr}
\end{equation}

\noindent \citep{wb00}. For the SIS, the shear is given by
\begin{equation}
\gamma_{\rm SIS}^{\phantom2}(x)={1\over2|x|}
\end{equation}

\noindent (SEF, page 244). For the Schwarzschild lens, the shear is
undefined at $x=0$ because of the singularity in 
$\kappa$ (eq.~[\ref{sigmasch}]). Otherwise,
\begin{equation}
\gamma_{\rm Sch}^{\phantom2}(x)={1\over x^2}
\end{equation}

Figures~\ref{shear05} and \ref{shear10}
show $\gamma_1$ and $\gamma_3$ (the shear of images~1 
and 3) versus source position~$\eta$ for halos located at
redhsift $z_L=0.5$ and 1 with various values of 
$M_{200}$, and sources located at redshift $z_S=3$ 
(notice that $\gamma_2$ is much less interesting, since image 2 is
usually very faint).
When a pair of curves appear (two curves with the same line
type in the same panel), the upper curve corresponds to $\gamma_1$ and
the lower one to $\gamma_3$, as the shear is always larger for the image
located closest to the lens. When only one curve appears, it
corresponds to $\gamma_3$. 

The Schwarzschild lens always has the largest shear, followed by the
SIS. The shear for the TIS is negligible at low masses, but becomes
more important as the mass increases, and at 
$M_{200}=10^{16}M_\odot$ it exceeds the shear for the NFW profile
and is comparable to the shear for the SIS.

In the limit $\eta\rightarrow0$, the shear converges to the value
$\gamma_1$, $\gamma_3=1$ for the Schwarzschild lens and 0.5 for the
SIS. The values for the NFW profile and the TIS vary with mass.
In the case of the TIS, the shear actually drops to 0 for low masses,
when only one image forms
(see case $M_{200}=10^{15}M_\odot$, $z_L=1.0$ in Fig.~\ref{shear05}). In this
case, a circular source near $\eta=0$ produces a nearly-circular image
near $\xi=0$. For larger masses, the same source produces three images,
with two of them (images 1 and 3) 
located near the tangential critical
circle, away for the center, where the shear is
large.  The Schwarzschild lens, SIS, and NFW profiles always reach this
limit at small enough $\eta$, for any value of $M_{200}$.

\subsection{Time Delay}

\begin{figure*}
\vspace{-20pt}
\includegraphics[width=156mm]{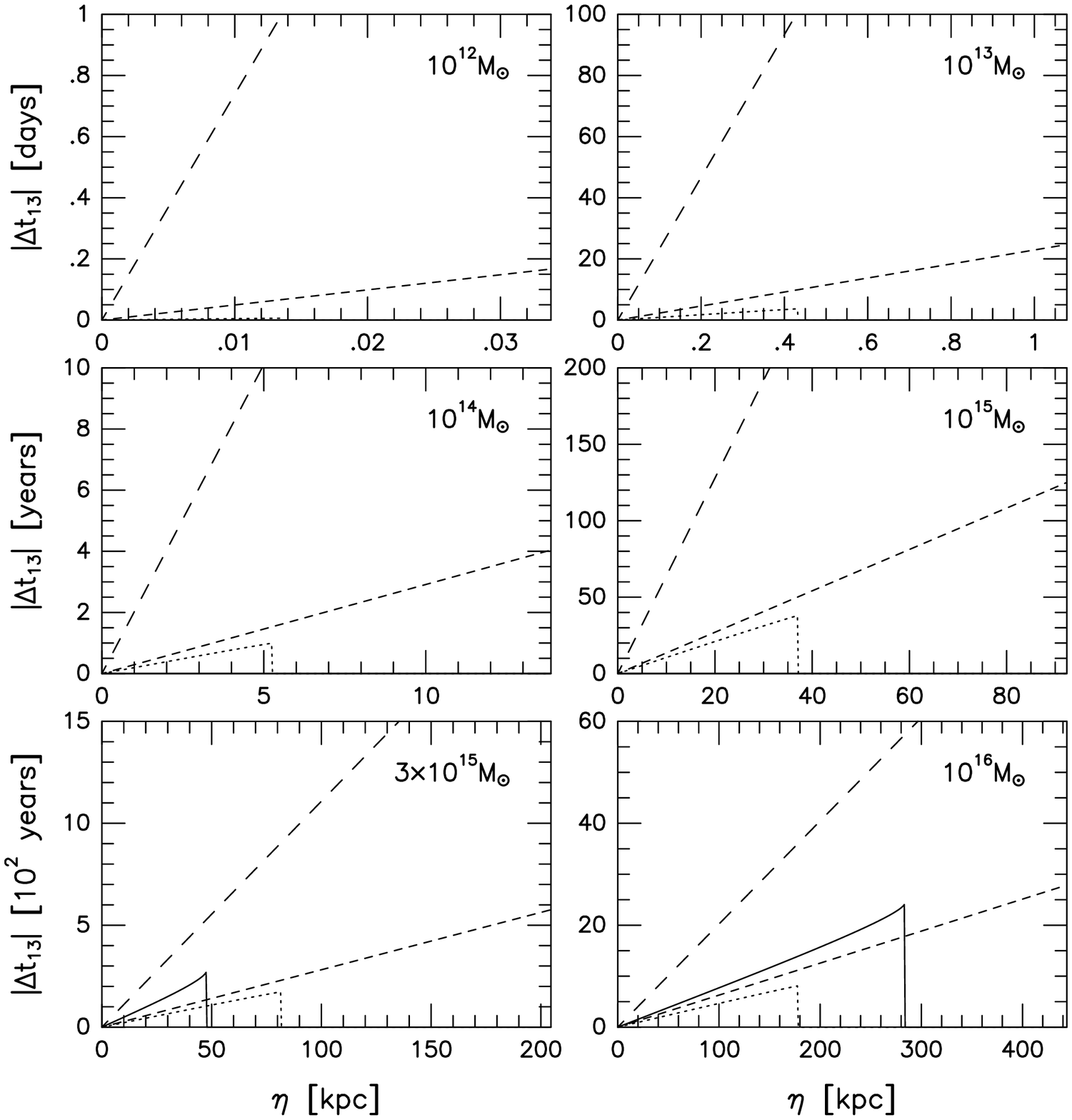}
 \vspace{-80pt}
 \caption{Time delay $|\tau_{13}|$ between images 1 and 3
versus source location $\eta$
for lenses of masses $10^{12}-10^{16}M_\odot$ located at $z_L=0.5$, for the
TIS (solid curves), the NFW profile (dotted curves), the SIS 
(short dashed curves) and the Schwarzschild lens (long dashed curves).}
\label{delay05}
\end{figure*}

\begin{figure*}
\vspace{-20pt}
\includegraphics[width=156mm]{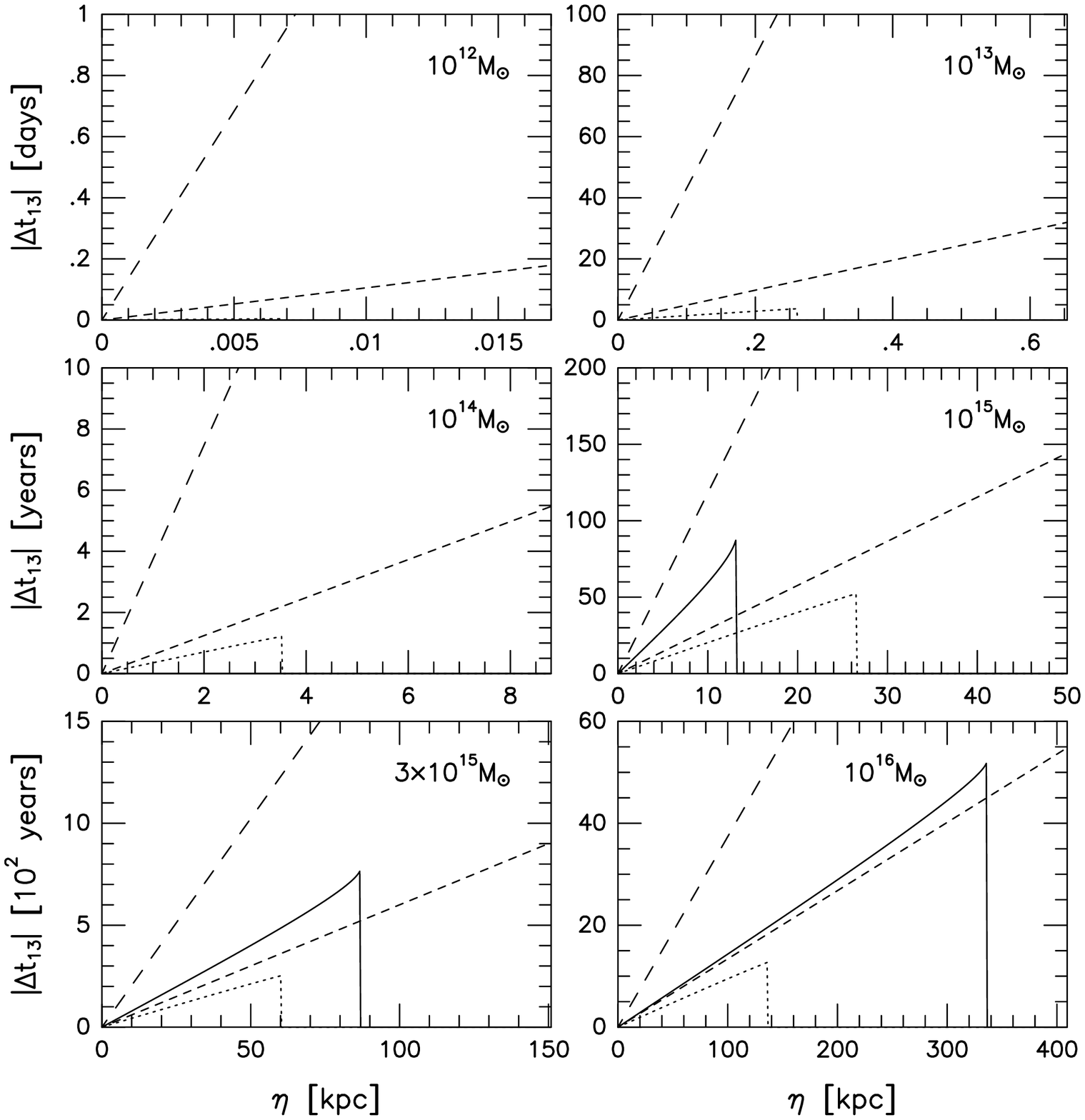}
 \vspace{-80pt}
 \caption{Time delay $|\tau_{13}|$ between images 1 and 3
versus source location $\eta$
for lenses of masses $10^{12}-10^{16}M_\odot$ located at $z_L=1.0$, for the
TIS (solid curves), the NFW profile (dotted curves), the SIS 
(short dashed curves) and the Schwarzschild lens (long dashed curves).}
\label{delay10}
\end{figure*}

For axially symmetric lenses, the deflection potential $\psi(x)$ is
defined by 
\begin{equation}
\label{potential}
\alpha={d\psi\over dx}\,.
\end{equation}

\noindent
For a source at location $y$ producing an image at location $x$, the
Fermat potential is defined by
\begin{equation}
\label{fermat}
\phi(x,y)={1\over2}(x-y)^2-\psi(x)\,,
\end{equation}

\noindent and the time delay $\Delta t_{ij}$
between two images located at $x_i$
and $x_j$, of a source located at $y$, is given by
\begin{equation}
\label{delay}
\Delta t_{ij}(y)={\xi_0^2D_S\over cD_LD_{LS}}
(1+z_L)\big[\phi(x_i,y)-\phi(x_j,y)\big]
\end{equation}

\noindent (SEF, eq.~[5.44]).
Hence, it is trivial to compute the time delay once the deflection
potential is known. For the TIS, we integrate equation~(\ref{alpha3}), 
and get
\begin{eqnarray}
\label{psitis}
\psi_{\rm TIS}^{\phantom2}(x)\!\!\!\!&=&\!\!\!\!{2ab\kappa_c\over Ab-Ba}\bigg\{
A\sqrt{a^2+x^2}-B\sqrt{b^2+x^2}\nonumber\\
&&\kern-40pt-Aa\ln\left[a+\sqrt{a^2+x^2}\right]
+Bb\ln\left[b+\sqrt{b^2+x^2}\right]\!\bigg\}.
\end{eqnarray}

\noindent For the NFW profile, we integrate
equation~(\ref{alphanfw}), and get
\begin{eqnarray}
\label{psinfw}
\psi_{\rm NFW}^{\phantom2}(x)\!\!\!\!&=&\!\!\!\!4\kappa_s\nonumber\\
&&\kern-40pt\times\cases{\displaystyle
{1\over2}\ln^2{|x|\over2}-2\arg\tanh^2\sqrt{1-|x|\over1+|x|}\,,
&$|x|<1\,;$\cr\noalign{\vspace{5pt}}
\displaystyle
{1\over2}\ln^2{|x|\over2}+2\arctan^2\sqrt{|x|-1\over|x|+1}\,,
&$|x|>1\,;$\cr
}
\end{eqnarray}

\noindent \citep{mbm03a,mbm03b}.
For the SIS, the deflection potential is given by
\begin{equation}
\psi_{\rm SIS}^{\phantom2}(x)=|x|\,,
\end{equation}

\noindent and for the Schwarzschild lens,
\begin{equation}
\psi_{\rm Sch}^{\phantom2}(x)=\ln|x|\,.
\end{equation}

\noindent Notice that for general lenses one normally computes the
deflection potential $\psi$ first, and then differentiates it to get the
deflection angle $\alpha$, not the other way around. It is the simplicity
of spherically symmetric lenses that enables us to compute $\alpha$
directly using equation~(\ref{alpha1}).

The time delay between images 1 and 3 (the two outermost images) is plotted
in Figures~\ref{delay05} and \ref{delay10},
for the same range of masses and source location,
lens redshift and source redshift as
in Figure~\ref{mag05}--\ref{shear10}. 
Unlike the magnification and shear, the time delay is
not dimensionless, and its value varies tremendously over the ranges
of $M_{200}$ and $\eta$ considered, going from hours to millenia.
For the SIS, the time delay varies linearly with $\eta$ (SEF, eq.~[8.36]).
For the Schwarzschild lens, the relation is more complicated
(SEF, eq.~[8.30]), but reduces to linear at small $\eta$.

For the ranges of $M_{200}$ and $\eta$ considered, the Schwarzschild lens
always produces the largest time delay, while the other three profiles produce
similar delays with the NFW profile always producing the
smallest delay. At low masses, the TIS produces one image, and therefore
no time delay, but
large masses it produces a larger time delay
than either the NFW profile or the SIS. This indicates that there is no
simple correspondence between the time delay and the central slope of
the density profile.

\section{WEAK LENSING}

\begin{figure}
\vspace{-20pt}
\hspace{-30pt}
\includegraphics[width=110mm]{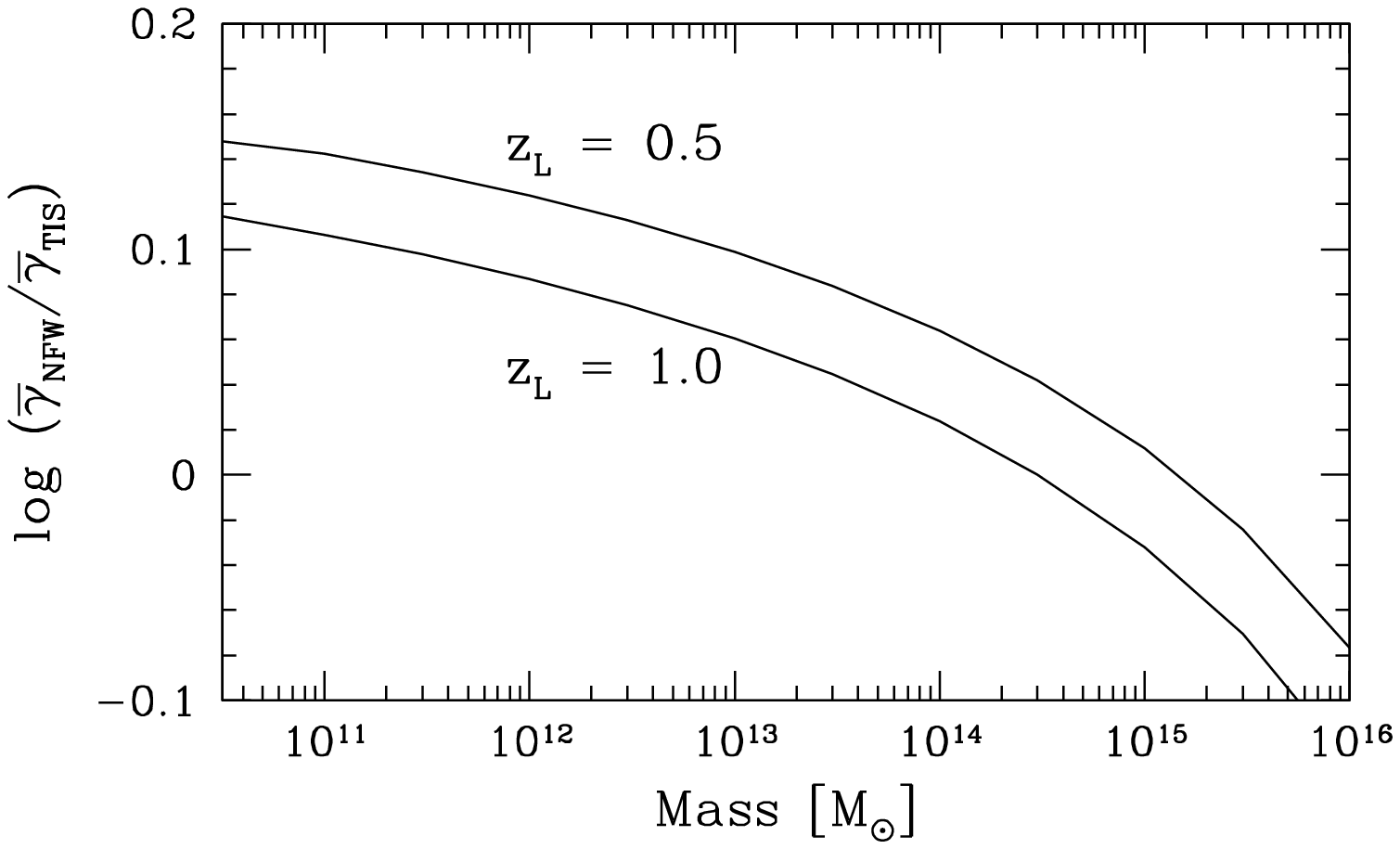}
 \vspace{-150pt}
 \caption{Ratio of the average shear inside radius $r_{200}$
for the NFW profile and the TIS profile, versus mass of the halo,
for halos located at redshifts $z_L=0.5$ and 1, as indicated.}
\label{weak}
\end{figure}

Weak lensing usually refers to the magnification and distortion of the
image of a background source by a foreground lens. Unlike strong lensing,
weak lensing normally does not produce multiple images of single sources.
The detection of coherent distortion patterns in the sky has been used
to constrain the mass of clusters. The first detections were reported by
\citet{twv90,bmf94,dml94,fahlman94},
and \citet{smail95}, followed by many others (see \citealt{bs01},
and references therein). More recently, the distortion pattern produced
by individual galaxies has also been detected
\citep{bbs96,griffiths96,dt96,ebbels98,hudson98,natarajan98,fischer00}.

Following the approach of \citet{wb00}, we use as measure
of the distortion produced by a lens the average shear $\bar\gamma$ inside
a distance $\xi=r_{200}$ from the lens center. In practice, this quantity
would be evaluated by averaging the shear of all images observed inside
$r_{200}$, after having eliminated foreground sources. We estimate this
quantity by integrating the shear over the projected area of the cluster.
The average shear inside radius $x$ is given by
\begin{eqnarray}
\label{shearavg1}
\bar\gamma(x)\!\!\!\!&=&\!\!\!\!{2\over x^2}\int_0^xx'\gamma(x')dx'
\nonumber\\
&&\qquad={2\over x^2}\int_0^xx'\left[{m(x')\over{x'}^2}-\kappa(x')\right]dx'\,.
\end{eqnarray}

\noindent We eliminate $m(x')$ using equation~(\ref{mx}). The two terms in the
integral can easily be computed using equations~(\ref{potential}) and
(\ref{alpha1}), respectively. Equation~(\ref{shearavg1}) reduces to
\begin{equation}
\label{shearavg2}
\bar\gamma(x)={2\over x^2}\left[\psi(x)-\psi(0)\right]-{\alpha(x)\over x}\,.
\end{equation}

For the Schwarzschild lens, $\psi(x)=\ln|x|$, and therefore $\bar\gamma$
diverges because of the term $\psi(0)$ in equation~(\ref{shearavg2}).
For the SIS, $\alpha(x)=x/|x|$, $\psi(x)=|x|$, and equation~(\ref{shearavg2})
reduces to
\begin{equation}
\label{shearavgsis}
\bar\gamma_{\rm SIS}^{\phantom2}(x)={1\over x}\,.
\end{equation}

\noindent We evaluate this expression at $r=r_{200}$, and get
\begin{equation}
\label{shearavgsis2}
\bar\gamma_{\rm SIS}^{\phantom2}
(r_{200})={400\pi\rho_cr_{200}\over3\Sigma_{\rm crit}}\,,
\end{equation}

\noindent where we used equation~(\ref{xscale}) to eliminate $x_{200}$, then
equation~(\ref{xiosis}) to eliminate $\xi_0$, and finally
equation~(\ref{sigmav}) to eliminate $\sigma_V^{\phantom2}$.
For the TIS, we substitute equations~(\ref{alpha3}) and
(\ref{psitis}) in equation~(\ref{shearavg2}), and get
\begin{eqnarray}
\label{shearavgtis}
\bar\gamma_{\rm TIS}^{\phantom2}(x)
\!\!\!\!&=&\!\!\!\!{2ab\kappa_c\over(Ab-Ba)x^2}\nonumber\\
&&\kern-35pt\times\Bigg\{
Aa\Bigg[\sqrt{1+{x^2\over a^2}}-1-2\ln{1+\sqrt{1+x^2/a^2}\over2}\Bigg]
\nonumber\\
&&\kern-35pt
-Bb\left[\sqrt{1+{x^2\over b^2}}-1
-2\ln{1+\sqrt{1+x^2/b^2}\over2}\right]\!\Bigg\}\,.
\end{eqnarray}

\noindent
We evaluate this expression at $r=r_{200}$, or equivalently 
$x=r_{200}/r_0\equiv\eta_{200}=24.2$. It reduces to
\begin{equation}
\label{shearavgtis2}
\bar\gamma_{\rm TIS}^{\phantom2}(r_{200})
=408.67{\rho_cr_{200}\over\Sigma_{\rm crit}}\,,
\end{equation}

\noindent where we used equation~(\ref{rho0}) to eliminate $\rho_0$.
Both $\bar\gamma_{\rm SIS}^{\phantom2}$ and 
$\bar\gamma_{\rm TIS}^{\phantom2}$ 
scale like $r_{200}$, or $M_{200}^{1/3}$. Their
ratio is given by
\begin{equation}
\left({\bar\gamma_{\rm SIS}^{\phantom2}
\over\bar\gamma_{\rm TIS}^{\phantom2}}\right)(r=r_{200})
=1.025\,,
\end{equation}

\noindent independent of the redshift $z_L$ and mass $M_{200}$ of the
lens. Furthermore, this
result is also independent of our assumptions that the background model
is $\Lambda$CDM, or the source is located at redshift $z_S=3$. These
enter only in the calculation of $\Sigma_{\rm crit}$ and $\rho_c$, 
which cancel out
when we take the ratio of equations~(\ref{shearavgsis2})
and (\ref{shearavgtis2}).
For the NFW profile, we substitute equations~(\ref{alphanfw}) and 
(\ref{psinfw}) in equation~(\ref{shearavg2}), and get
\begin{eqnarray}
\bar\gamma_{\rm NFW}^{\phantom2}\!\!\!\!&=&\!\!\!\!
{4\kappa_s\over x^2}\Bigg[\ln^2{x\over2}-\ln{x\over2}
-4\arg\tanh^2\sqrt{1-x\over1+x}\nonumber\\
&&\qquad\qquad-{2\over\sqrt{1-x^2}}\arg\tanh\sqrt{1-x\over1+x}\,\Bigg]\,.
\end{eqnarray}

\noindent To evaluate this expression at $r_{200}$, we simply set
$x=r_{200}/r_{\rm NFW}^{\phantom2}=1/c$.

In Figure~\ref{weak}, we plot the ratio 
$\bar\gamma_{\rm NFW}^{\phantom2}/\bar\gamma_{\rm TIS}^{\phantom2}$
versus mass, for lenses located at redshifts $z_L=0.5$ and 1,
and sources located at redshift $z_S=3$. The ratio
$\bar\gamma_{\rm NFW}^{\phantom2}/\bar\gamma_{\rm SIS}^{\phantom2}$
can be obtained by shifting the curves down
by an amount $\log1.025=0.0107$. The ratios are larger at lower
redshifts $z_L$. They also decrease with increasing mass. This reflects the
fact that as the mass increases, the concentration parameter $c$ of the NFW
profile decreases, while the ratio $r_{200}/r_0$ for the TIS remains
fixed at 24.2. This figure can be {\it qualitatively} compared with the
top left panel of Figure~3 in \citet{wb00}, after the
appropriate shifting. Notice, however, that (1) the $\Lambda$CDM used by
these authors is not exactly the same as the one we use, and (2) these
authors used the approach of \citet{nfw97} to compute the
concentration parameter, while we use the more recent approach of
Eke et al. (2001). At large enough masses, the average shear produced
by the TIS exceeds the one produced by the NFW profile.

We reach essentially the same conclusion as \citet{wb00},
namely that using the average shear to estimate the mass of lensing halos
can lead to considerable errors if the wrong density profile is
assumed. At small masses, 
$\bar\gamma_{\rm NFW}^{\phantom2}/\bar\gamma_{\rm TIS}^{\phantom2}>1$,
and therefore the mass of a TIS would be underestimated if the lens
is incorrectly assumed to follow a NFW profile. At high mass, the
true mass of a TIS would be overestimated.

\section{DISCUSSION}

In the previous two sections, we derived the effects of strong and 
weak lensing, respectively. Here, ``strong lensing'' refers to case
with multiple images, arcs, or rings, while ``weak lensing'' refers
to the magnification and shear of single images. We can divide the
observed cases of strong lensing in two groups. The first group contains 
the ``arc second'' cases: multiple-image systems with image separations
of order arc seconds, or rings with radii of that order
(see \citealt{kochanek98}).\footnote{See http://cfa-www.harvard.edu/castles.} 
In most cases, the lens is a
single, massive galaxy, with possibly some additional
contribution from the environment in which this galaxy is located
(\citealt{tog84}; see, however, \citealt{pm03}).
A classic example is Q0957+561, the first gravitational lens
to be discovered. 
The second group contains the ``arc minute'' cases, in which the lens
is an entire cluster of galaxies. These lenses produce mostly giant arcs,
with radii in the range $15''-60''$ (see Table~1 of \citealt{wnb99}). 
The most famous case is the cluster
CL~0024+1654, which produces multiple arcs.

We showed that for all profiles considered, the image separation is weakly
dependent on the source location (Fig. \ref{separation}),
when multiple images actually
form. If we neglect this dependence, the image separation
$\Delta\theta\approx2\theta_t$ can be read off the top panels of 
Figure~\ref{critfig2}.
We see immediately that galaxy-size objects cannot produce
arc-second separations if they are described by the TIS or the NFW profile
(under the assumptions described in \S2 that $z_{\rm coll}=z_L$ for TIS halos
while $c$ for NFW halos is the typical value for halos at $z_{\rm obs}=z_L$).
The TIS does not produce multiple images unless the mass is of order of
a cluster mass or above. The NFW profile can produce multiple images for
any mass, but a separation $\Delta\theta>1''$ requires a mass
of order $10^{13}M_\odot$. Even the SIS needs a mass in excess of
$10^{12}M_\odot$ to produce arc-second separations.  
At the cluster scale, all profiles are capable of producing arc-minute
separations. In this limit, for a given mass, the separation is larger for
the Schwarzschild lens, smaller for the NFW profile (because of the
small concentration parameter at large mass), and comparable
for the SIS and TIS. As we indicated in
the introduction, the TIS and NFW profile are applicable to dwarf
galaxies and clusters of galaxies, but might not be applicable to 
ordinary
galaxy-scale objects because baryonic processes are neglected. Hence, the
inability of the TIS and NFW profiles to produce arc seconds separations
with galaxy-size lenses is not a concern.

There is very important caveat to this conclusion. So far, we have assumed
that the redshift $z_L$ where the lensing halo is located is the same as
the redshift $z_{\rm coll}$ at which this halo first collapsed and virialized.
As we pointed out in \S2,
this does not have to be the case, however, since in principle the halo could have collapsed at any redshift $z_{\rm coll}\geq z_L$.
Consider for instance the TIS. The central density $\rho_0$ is determined by
equation~(\ref{rho0}), and so far we have assumed that $\rho_c(z)$ was
evaluated at $z_L$. But if $z_{\rm coll}>z_L$, then $\rho_c(z)$ must be
evaluated at $z_{\rm coll}$, instead. 
By combining
equations~(\ref{m200}), (\ref{r0}), (\ref{rho0}), and~(\ref{kappac}),
and get
\begin{equation}
\label{kappacoll}
\kappa_c\propto{\rho_0r_0\over\Sigma_{\rm crit}(z_L)}
\propto{M_{200}^{1/3}\rho_c^{2/3}(z_{\rm coll})\over\Sigma_{\rm crit}(z_L)}\,.
\end{equation}

\noindent Hence, for a given mass, increasing $z_{\rm coll}$
will result in an increase in $\kappa_c$,
and lensing will be stronger.
Similarly, a particular effect (for instance, production of multiple
images, which requires $\kappa_c>1$) 
can be achieved with a smaller mass halo if that halo collapsed
earlier. For the NFW profile, the expressions for
the characteristic radius and density, equations~(\ref{rnfw})
and~(\ref{rhonfw}), depend on the concentration parameter $c$. The
models that describe the dependence of $c$ on $M_{200}$, and $z$ 
\citep{nfw97,ens01} already take into account the
difference between $z$ and $z_{\rm coll}$. Hence, unlike the case for the TIS,
we are not free to set the redshift $z$ to any other value than
$z_L$ in equation~(\ref{rhonfw}). However, as we explained in \S2, the
value of $c$ we have used so far, which is a single function of
$M_{200}$ and $z_L$, is a statistical average over many realizations,
and a particular halo might have a different concentration parameter.

Let us now focus on the TIS, and perform a simple calculation to
estimate the probability that a TIS can produce multiple images.
To compute $\kappa_c$, we substitute
equations~(\ref{m200}), (\ref{r0}), and (\ref{rho0}) in
equation~(\ref{kappac}). For the particular cosmological model ($\Lambda$CDM)
and redshifts ($z_S=3$, $z_L=0.5$, 1.0) we have considered in this paper,
we get
\begin{eqnarray}
\label{kappacoll2}
\kappa_c\!\!\!\!&=&\!\!\!\!480.75{\rho_c^{2/3}(z_L)\over\Sigma_{\rm crit}(z_L)}
\left[{\rho_c(z_{\rm coll})\over\rho_c(z_L)}\right]^{2/3}M_{200}^{1/3}
\nonumber\\
&&=\left[{\rho_c(z_{\rm coll})\over\rho_c(z_L)}\right]^{2/3}M_{15}^{1/3}\times
\cases{0.968\,,&$z_L=0.5\,;$\cr
1.289\,,&$z_L=1.0\,;$\cr}
\end{eqnarray}

\noindent where $M_{15}\equiv M_{200}/10^{15}M_\odot$. 
Strong lensing requires $\kappa_c>1$. If $z_L=z_{\rm coll}$, this condition
becomes
\begin{equation}
M_{15}\geq\cases{1.102\,,&$z_L=0.5\,;$\cr
0.467\,,&$z_L=1.0$\cr}
\end{equation}

\noindent (these values are the location of the cutoffs in Fig.~\ref{critfig2}).
For the cosmological model we consider, a 1-$\sigma$ density
fluctuation collapsing at redshift $z_{\rm coll}=0.5$ (1.0) has a mass of
about $M_{15}=2\times10^{-3}$
($4\times10^{-4}$).
Such ``typical'' objects will not be capable of
producing multiple images of a source at redshift $z_S=3$, since the
resulting value $\kappa_c=0.122$ (0.095) is smaller than unity. 
This simply indicates that
multiple images are not produced by  the typical objects that collapse
at $z_{\rm coll}=z_L$,
which is certainly consistent with the fact that fewer than 100 multiple-image
systems have been observed. 

Increasing $\kappa_c$ above
unity would require an object about a thousand times
more massive than a typical object at the same redshift. Objects of this
mass are rare but do exist. We can make a simple estimate of how atypical
such a massive object is. Over most of the mass range of cosmological
interest (from small
galaxies to clusters of galaxies) the CDM power spectrum can
be roughly approximated
by a power law $P(k)\propto k^n$, where $k$ is the wavenumber
and $n\approx-2$. The rms density 
fluctuation $\delta_{\rm rms}$ is then given by 
$\delta_{\rm rms}\approx k^{3/2}P^{1/2}(k)\propto k^{1/2}$.
At a given redshift, different values of
the wavenumber $k$ correspond to different mass scales 
$M$ according to $M\propto k^{-3}$. The relation between rms
density fluctuation and mass scale at fixed epoch is therefore approximated by
\begin{equation}
\delta_{\rm rms}\propto M^{-1/6}\,.
\end{equation}

\noindent Increasing the mass by a factor of 1000 therefore reduces
$\delta_{\rm rms}$ 
by a factor of $1000^{1/6}\approx3$. Because of the reduction
in $\delta_{\rm rms}$, 
a 1-$\sigma$ fluctuation ($\delta=\delta_{\rm rms}$) at this higher mass
will no longer collapse by the same redshift (it will collapse later),
but a 3-$\sigma$ fluctuation ($\delta=3\delta_{\rm rms}$)
will. Such fluctuations are rare, but not
vanishingly rare. In Gaussian statistics, the probability that a randomly
located point in space is inside a 3-$\sigma$ density fluctuation
(i.e. $\delta\geq3\delta_{\rm rms}$) is 1/384.
Hence, one of every 384 halos would be capable of producing multiple images
(of course, whether any halo actually produces multiple images
depends on the location of the sources). This was derived by assuming
$z_{\rm coll}=z_L$. As equation~(\ref{kappacoll2}) shows, the condition
$\kappa_c>1$ can be satisfied with a smaller mass it we assume that the
halo collapsed at an earlier redshift $z_{\rm coll}>z_L$.

A similar calculation could be carried out for the NFW profile. In this
case, at fixed mass $M_{200}$ and redshift
$z_L$, we are free to choose a concentration
parameter $c$ that differs from the statistical average corresponding to this
mass and redshift. If we then combine equations~(\ref{rnfw}), (\ref{rhonfw}),
and (\ref{kappas}), and ignore the weak dependence of the denominator
of equation~(\ref{rhonfw}) on $c$ at large $c$, we get
\begin{equation}
\label{kappascoll}
\kappa_s={\rho^{\phantom2}_{\rm NFW}r^{\phantom2}_{\rm NFW}
\over\Sigma_{\rm crit}(z_L)}
\propto{c^2M_{200}^{1/3}\rho_c^{2/3}(z_L)\over\Sigma_{\rm crit}(z_L)}\,.
\end{equation}

\noindent Hence, we can increase the effect of lensing, or reduce the
minimum mass necessary to produce a particular effect, by increasing the
concentration parameter. 

The simple calculations presented in this discussion were for
illustration purpose only. In a future paper, we will present a detailed
calculation of the expected frequency of multiple image systems for
comparison with the statistics of observed lensing.

\section{SUMMARY AND CONCLUSION}

We have derived the lensing properties of cosmological halos
described by the Truncated Isothermal Sphere model. The solutions depend
on the background cosmological model through the critical surface
density $\Sigma_{\rm crit}$, which is a function of the cosmological
parameters and the source and lens redshifts, and the TIS parameters
$\rho_0$ and $r_0$, which are functions of the mass and collapse redshift
of the halo, and the cosmological parameters. By expressing the surface
density of the halo in units of $\Sigma_{\rm crit}$ and the distances
in units of $r_0$, all explicit dependences on the cosmological
model disappear, and the solutions are entirely expressible in terms of two
dimensionless parameters, the central convergence $\kappa_c$ and
the scaled position $y$ of the source. We have computed
solutions for the critical curves and caustics, the image separations, the
magnification and brightness ratios, the shear, and the time delay.
The ability of the TIS to produce strong lensing (multiple
images and rings) depends entirely on $\kappa_c$. If $\kappa_c<1$,
only one image can form. If $\kappa_c>1$,
either one or three images can form, depending on
whether the source is located outside or inside the radial caustic.
When three images are produced, the central one is usually very faint, being
highly demagnified. Degenerate image configurations occur when an
extended source overlaps a caustic. Two images are produced when the
source overlaps the radial caustic, while an Einstein ring with a
central spot is produced when the source overlaps the tangential caustic,
which is a single point located at $y_t=0$.
These degenerate cases correspond to maxima of the total
magnification, which diverges as the source size goes to zero. When
three images are produced, the angular separation between the
two outermost images depends strongly on $\kappa_c$, but only
weakly on the source location.

For comparison, we derived (or extracted from the literature) the
lensing properties of three comparison models: the NFW profile,
the singular isothermal sphere, and the Schwarzschild lens.
Unlike the TIS, all of these profiles have a central singularity,
which allows them to produce multiple images at any mass, provided
that the source is sufficiently aligned with the lens. In practice,
image separations large enough to be resolved can be achieved by
galactic-mass objects only for the Schwarzschild lens, and by
supergalactic-mass objects for all profiles.

This paper focused on the intrinsic properties of individual lenses
described by the TIS model and the comparison models. 
It provides all the necessary formulae
one needs to study gravitational lensing by the TIS, and 
the comparison profiles, in specific
cosmological models. We will present such studies in forthcoming papers.
As an illustration here, we applied the TIS model to the currently-favored
$\Lambda$CDM universe, to calculate the central convergence $\kappa_c$
expected for TIS halos of different masses and collapse epochs.
We found that high-redshift sources (e.g. $z_S\approx3$) will be
strongly lensed by TIS halos (i.e. $\kappa_c>1$) only for cluster-mass halos,
assuming that these halos formed at the redshift they are observed. 
As equation~(\ref{kappacoll2}) shows, the halo mass required for
strong lensing can be decreased by increasing the formation redshift
of the halo. An example of a lens for which $z_{\rm coll}>z_L$ is
discussed by \citet{si00}, who showed that the
mass-model derived by \citet{tkd98} to explain their lensing data for
cluster CL 0024+1654 at $z=0.39$ is very well-fit by a TIS halo with
$\rho_0\approx0.064h^2M_\odot\,{\rm pc}^{-3}$ and
$r_0\approx20h^{-1}{\rm kpc}$. This central density implies that
the halo collapse redshift is $z_{\rm coll}\approx2.5$ (i.e.
$z_{\rm coll}\gg z_L$).

\section*{Acknowledgments}

We thank J. Navarro for providing the FORTRAN subroutine {\tt cens.f}
for computing
the concentration parameter for the NFW profile.
This work was supported by NASA ATP Grants NAG5-10825 and NAG5-10826,
and Texas Advanced Research Program
Grant 3658-0624-1999.

%

\bsp

\label{lastpage}

\end{document}